\documentclass[letterpaper, 10 pt, conference]{ieeeconf}

\IEEEoverridecommandlockouts                              
\usepackage{amsmath,amsfonts,amssymb}
\usepackage{graphicx,subfigure}

\def \V {{\mathcal V} }

\def \tildex {\tilde{x}}

\newfont{\gothic}{eufm10 scaled\magstep0}

\newcommand{\rr}{\mbox{$\mathbb R$}}
\newcommand{\nn}{\mbox{$\mathbb N$}}

\newtheorem{theorem}{Theorem}[section]
\newtheorem{lemma}[theorem]{Lemma}
\newtheorem{corollary}[theorem]{Corollary}

\newtheorem{definition}[theorem]{Definition}
\newtheorem{make_remark}[theorem]{Remark}

\begin{document}

\title{Riccati observers for position and velocity bias estimation\\from either direction or range measurements}

\author{Tarek~Hamel and Claude Samson
\thanks{T. Hamel is with I3S UNS-CNRS, Nice-Sophia Antipolis, France, {\tt thamel@i3s.unice.fr}.}
\thanks{C. Samson is with INRIA and  I3S UNS-CNRS, Sophia Antipolis, France, {\tt claude.samson@inria.fr, csamson@i3s.unice.fr}. }}

\maketitle

\begin{abstract}
This paper revisits the problems of estimating the position of an object moving in $n$ ($\geq 2$)-dimensional Euclidean space using velocity measurements and either direction or range measurements of one or multiple source points. The proposed solutions exploit the Continuous Riccati Equation (CRE) to calculate observer gains yielding global exponential stability of zero estimation errors, even in the case where the measured velocity is biased by an unknown constant perturbation. These results are obtained under persistent excitation (p.e.) conditions depending on the number of source points and body motion that ensure both uniform observability and good conditioning of the CRE solutions.  With respect to prior contributions on these subjects some of the proposed solutions are entirely novel while others are adapted from existing ones with the preoccupation of stating simpler and more explicit conditions under which uniform exponential stability is achieved. 
A complementary contribution, related to the delicate tuning of the observers gains, is the derivation of a lower-bound of the exponential rate of convergence specified as a function of the amount of persistent excitation. Simulation results illustrate the performance of the proposed observers.

\end{abstract}
~
\section{Introduction}
The general problem of estimating the position, or the complete pose (position and orientation), of a body relatively to a certain spatial frame is central for a multitude of applications. This is common knowledge. Among all sensing modalities that can be used to acquire the necessary information, source points direction (or bearing) measurements has early motivated many studies, in particular for pose estimation when body and source points are motionless in the frame of interest, a problem referred to as the Perspective-$n$-Point (PnP) problem in the dedicated literature \cite{Haralick94}. Proposed solutions may roughly be classified into two categories, namely {\em non-iterative} methods based on a finite set of measurements (one per source point) feeding polynomial equations that are either algebraically or numerically solved \cite{kneip2011novel}, and {\em iterative} methods involving ongoing measurements that feed gradient-like recursive algorithms \cite{Haralicketal1989,Janabi2010}. Such recursive algorithms are called {\em observers} in the Automatic Control community. Generically, at least three source points are needed to ensure that the number of body poses compatible with the measurements is finite \cite{Haralick94}. It is commonly acknowledged that these two types of methods are complementary. Non-iterative methods are of interest to work out an approximation of the body pose after elimination of non-physical solutions, whereas iterative methods, that are local by nature (since they may be stuck at local minima even in the case of a unique global minimum), allow for a more precise estimation in relation to their filtering properties \cite{1996_Soatto_TAC}.
The present paper focuses on the sole estimation of the body position. This corresponds to applications for which the body's attitude is either of lesser importance or is estimated by using other sensing modalities. In this case, iterative methods are all the more interesting that their domain of convergence can be global. The reason is that, without the compact group of rotations being involved, this simplified problem is amenable to exact linearisation and can be associated with globally convex cost functions, as shown further in the paper. Another advantage of iterative methods is that they are naturally suited to handle the non-static case, i.e. when either the body or the point source(s) move(s), by using on-line the extra data and information resulting from motion. In particular, the observation of a single point source may be sufficient in this case, provided that the body motion regularly grants a sufficient amount of "observability".
This possibility has been studied recently in \cite{Batista2013} where the problem is linearised by considering an augmented state vector. Another solution, not resorting on state augmentation, is proposed in \cite{LebHamMahSam15}. The present paper offers a generalization of previous studies on this subject that encompasses the static and non-static cases with an arbitrary number of source points.

{\em Global Navigation Satellite Systems} (GNSS), and the American {\em Global Positioning System} (GPS) \cite{Dixon91} in particular, have familiarized the larger public with the problem of body position estimation from source points distance (or range) measurements. In the static or quasi-static case, solutions to this problem may again be classified into non-iterative and iterative methods. Similarly to the direction measurements case, three point sources (satellites) are also required to obtain a finite number (equal to two) of theoretical solutions, with an extra source point (non-coplanar with the other points) needed to eliminate the non-physical solution and overcome the problem of desynchronized clocks resulting in constant range measurement bias. The resolution of this problem is also facilitated by the fact that constraint (output or measurement) equations can be made linear in the unknown position coordinates. Studies of the non-static case are much less numerous and more recent \cite{Batista14,Batista14a}. To our knowledge, Batista and al. \cite{Batista11} were first to address this case by exploiting the possibility of linearising the estimation problem via state augmentation, even when the body velocity vector is biased by a constant vector. A similar idea is used in the present paper, but via lower-dimensional state augmentation. This yields simpler observers and reduced computational weight.

For five decades, Kalman filters for linear systems, and their extensions to non-linear systems known as Extended Kalman Filters (EKF), have consistently grown in popularity near engineers with various backgrounds (signal processing, artificial vision, robotics,...) to address a multitude of iterative state estimation problems involving additive "noise" upon the state and/or the measurements. The optimality of these filters in a stochastic framework under specific noise conditions and assumptions, and their direct applicability to Linear Time-Varying (LTV) systems, have undoubtedly contributed to this popularity. It is however important to keep in mind, or to recall, that the stability and robustness properties associated with them, i.e. features that supersede conditional stochastic optimality in practice, are not related to stochastic issues. They result from properties of the associated deterministic continuous-time (or discrete-time, depending on the chosen computational framework) Riccati equation that underlies a (locally) convex estimation error index (or Lyapunov function) and a way of forming recursive estimation algorithms that uniformly decrease this index exponentially (under adequate observability conditions). With this perspective, Kalman filters belong to the (slightly) larger set of Riccati observers that we intentionally derive here in a deterministic framework, knowing that a complementary stochastic interpretation may be useful to subsequently tune the Riccati equation parameters and observer gains. We also believe that, by contrast with standard Kalman filter derivations, this approach allows one to better comprehend how the system observability properties are related to the good conditioning of the Riccati equation solutions and to the observer's performance (the rate of convergence to zero of the estimation errors, in particular) via a Lyapunov analysis.

The research themes addressed in the present paper are not new, nor are the basic conceptual tools (Riccati equation, Lyapunov stability, uniform observability and persistent excitation,...) used to derive the proposed observers. However, we believe that our approach to the problems and the resulting observer design synthesis are original. Also, by contrast with a majority of studies based on the application of Kalman filtering, uniform exponential stability of the observers is rigorously proved in association with explicit and simple observability conditions worked out from the corresponding observability Grammian condition. The connection between rate of convergence and amount of observability is also drawn out explicitly.
Observers are derived for both direction measurements and range measurements, in $n$ ($\geq 2$)-dimensional Euclidean space so that both 2D and 3D cases (of particular practical interest) are covered, with an arbitrary number of source points. Concerning this latter aspect, the observers are designed by first considering a single source point, with stability and convergence of the observer relying on persistent excitation properties associated with the body motion. The solutions are then generalized to the case of multiple source points, with the augmentation of the number of these points reflecting on the gradual weakening of the body motion conditions needed to ensure uniform exponential stability. While measuring the body velocity is central to estimate the position, we also show how to modify the observers via state augmentation when velocity measurement are biased by a constant vector. Uniform exponential stability is then preserved under either the same observability conditions, when direction measurements are used, or slightly stronger ones, when range measurements are used with less than $(n+1)$ source points. A complementary original result concerns the case of range measurements corrupted by an unknown common bias.



The paper is organized along six sections.  Following the present introduction, Section~\ref{sec:recalls} recalls basic observability concepts and central properties of the CRE, complemented with a few original technical results used for the design and analysis of the observers. Direction measurements and range measurements cases are treated in Sections~\ref{sec:direction} and \ref{sec:range} respectively. Illustrative simulations results are presented in Section~\ref{sec:simulations}, followed by a short section \ref{sec:conclusions} of concluding remarks. The proofs of several technical results are reported in the Appendix. 


\section{Recalls} \label{sec:recalls}
Although several of the definitions and results recalled in this section are well known, others are not. Our main intent here is to provide the reader with a self-contained overview of basic observability concepts and of state observers whose gains are calculated from solutions to the Continuous Riccati Equation (CRE). This overview is also an opportunity to recall natural Lyapunov functions associated with these observers for stability and convergence analysis.

Throughout the paper the following notation is used:
\begin{itemize}
\item $A(t)$, $B(t)$, $C(t)$ are finite-dimensional matrix-valued functions depending on time. They are continuous, bounded, and $r$ ($\geq 0$) times differentiable with bounded derivatives, with $r$ specified (sometimes implicitly) in subsequent developments.
\item The abbreviation p.s.d. (resp. p.d.) is used to denote semipositive (resp. positive) square matrices   that are also symmetric. For instance, square null matrices are p.s.d. matrices and identity matrices, denoted as $I_d$ independently of their dimensions, are p.d. matrices. The set of p.s.d. matrices obviously contains the set of p.d. matrices.
\item $Q(t)$ and $V(t)$ are p.s.d. finite-dimensional matrix-valued functions of time. They are also continuous and bounded. When no specific indication is provided in the text these matrix-valued functions are chosen strictly positive and greater than $\epsilon I_d$ with $\epsilon >0$.  
\item The infimum (resp. supremum) over time of the smallest (resp. largest) eigenvalue of a p.s.d matrix-valued function $P(t)$ is denoted as $p_m$ (resp. $p_M$). For the matrix-valued function $V(t)$ these infimum and supremum values are accordingly denoted as $v_m$ and $v_M$.
\end{itemize}

\subsection{Observability definitions and conditions} \label{observability}
Consider a generic linear time-varying (LTV) system
\begin{equation} \label{linear_system}
\left\{ \begin{array}{lll}
\dot{x} & = & A(t)x+B(t)u \\
y & = & C(t)x
\end{array} \right.
\end{equation}
with $x \in \rr^n$ the system state vector, $u \in \rr^s$ the system input vector, and $y \in \rr^m$ the system output vector. The following definitions and properties of observability associated with this system are borrowed from \cite{chen1984}.

\begin{definition}[instantaneous observability] \label{instantaneous_observability}
System \eqref{linear_system} is {\em instantaneously observable} if $\forall t$, $x(t)$ can be calculated from the input $u(t)$, the output $y(t)$, and the time-derivatives $u^{(k)}(t)$, $y^{(k)}(t)$, $k \in \nn$.
\end{definition}

\begin{lemma}
Define the observation space at the time-instant $t$ as the space generated by
\[
{\cal{O}}(t):=\left( \begin{array}{c}
N_0(t)\\
N_1(t)\\
\vdots
\end{array} \right)
\]
with $N_0=C$, $N_{k+1}=N_{k}A+\dot{N}_k$, $k=1,\ldots$ Then System \eqref{linear_system} is instantaneously observable if $rank({\cal{O}}(t))=n$.
\end{lemma}

\begin{definition}[uniform observability] \label{uniform_observability}
Sytem \eqref{linear_system} is {\em uniformly observable} if there exists $\tau>0$ such that, $\forall t$, $x(t)$ can be calculated from the knowledge of the input $u(.)$ and ouput $y(.)$ on the time-interval $[t,t+\tau]$.
\end{definition}
Note that, with this definition, uniformity is related to time and not to the input. This definition of uniform observability, which we adopt here, is thus different from other definitions proposed in the literature, {\em e.g.} \cite{gauthier-kupka-1994} or \cite{besancon-2007}.
\begin{theorem}[sufficient condition for uniform observability]  \label{chen1984}
Sytem \eqref{linear_system} is {\em uniformly observable} if there exist $\delta>0$, $\mu>0$ such that $\forall t\geq 0$
{\small
\begin{equation} \label{grammian}
W(t,t+\delta):=\frac{1}{\delta} \int_t^{t+\delta}\Phi^{\top}(s,t)C^{\top}(s)C(s)\Phi(s,t)ds \geq \mu I_d >0
\end{equation}
}
with $\Phi(t,s)$ the transition matrix associated with $A(t)$, i.e. such that $\frac{d}{dt}\Phi(t,s)=A(t)\Phi(t,s)$ with $\Phi(t,t)=I_d$.
\end{theorem}

The matrix valued-function $W(t,t+\delta)$ is called the {\em observability Grammian} of System \eqref{linear_system}.
A very useful result, derived in \cite{scandaroli-2013}, gives a sufficient condition for uniform observability in terms of the properties of the matrices $A(t)$ and $C(t)$ and their time-derivatives

\begin{lemma} \label{scandaroli}
If there exists a matrix-valued function $M(.)$ of dimension $(p \times n)$ ($p \geq 1$) composed of row vectors of $N_0(.)$, $N_1(.)$,$\hdots$, such that for some (strictly) positive numbers $(\bar{\delta},\bar{\mu})$ and $\forall t \geq 0$
\begin{equation} \label{M}
\frac{1}{\bar{\delta}} \int_t^{t+\bar{\delta}}M^{\top}(s)M(s)ds \geq \bar{\mu} I_d >0
\end{equation}
then the observability Grammian of System \eqref{linear_system} satisfies the condition \eqref{grammian} (and this system is thus uniformly observable).
\end{lemma}

\subsection{Riccati observers} \label{riccati-observers}
We here call {\em Riccati observer} any observer of System \eqref{linear_system} of the form
\begin{equation} \label{riccati-observer}
\dot{\hat{x}}=A(t)\hat{x}+B(t)u+K(t)(y-C(t)\hat{x})~;~\hat{x}(0) \in \rr^n
\end{equation}
with the observer gain given by
\begin{equation} \label{riccati-gain}
K(t)=k(t)P(t)C^{\top}(t)Q(t)~;~k(t)\geq 0.5
\end{equation}
where $P(t)$ is the solution to the so-called Continuous Riccati Equation (CRE)
\begin{equation} \label{dP}
\dot{P}=A(t)P+PA^{\top}(t)-PC^{\top}(t)Q(t)C(t)P+V(t)
\end{equation}
with $P(0)$ any p.d. matrix and $Q(t)$, $V(t)$ p.s.d. matrices that have to be specified.
Note that the optimal Kalman gain in the stochastic setting where the matrices $V(t)$ and $Q^{-1}(t)$ are p.d. matrices and interpreted as covariance matrices of additive noise on the system state and output is obtained by taking $k(t)=1$.

Let us now quickly recall how the stability and convergence properties of a Riccati observer is directly related to the properties of the solution $P(t)$ to the CRE. Define the estimation error $\tilde{x}:=x-\hat{x}$, from \eqref{linear_system} and \eqref{riccati-observer} one obtains the error equation
\begin{equation} \label{estimation-error}
\dot{\tilde{x}}=(A(t)-K(t)C(t))\tilde{x}
\end{equation}
Assume (for the time being) that $P(t)$, which is a symmetric matrix by construction, is well defined on $\rr^+$ and is p.d., so that its inverse is also well defined and p.d., and consider the candidate Lyapunov function $\V(t)=\tildex^{\top}(t)P^{-1}(t)\tildex(t)$. Then using the fact that the time-derivative of $\dot{P}^{-1}$ satisfies the equation
{\small
\begin{equation} \label{dPinverse}
\dot{P}^{-1}=-P^{-1}A(t)-A^{\top}(t)P^{-1}+C^{\top}(t)Q(t)C(t)-P^{-1}V(t)P^{-1}
\end{equation}
}
and using \eqref{riccati-gain} and \eqref{estimation-error}, one easily verifies that the time-derivative of $\V(t)$ is given by
{\small
\begin{equation} \label{dV}
\begin{array}{lll}
\dot{\V}&=&-\tildex^{\top} \big( (2k(t)-1)C^{\top}(t)Q(t)C(t)+P^{-1}V(t)P^{-1} \big) \tildex \\
~ & \leq & -\tildex^{\top} P^{-1}V(t)P^{-1} \tildex \\
~ & \leq & -\frac{p_m^2}{p_M}v_m \V  ~~~(\leq 0)
\end{array}
\end{equation}
}
so that $\V(t) \leq \V(0) exp(-\frac{p_m^2}{p_M}v_m t)$.
To conclude that $\tildex=0$ is globally exponentially stable it thus suffices to choose $V(t)>v_mI_d$ with $v_m>0$ and to show that $P(t)$ i) is always well-defined, ii) that it is p.d., and --most importantly-- iii) that it is well conditioned in the sense that $p_m$ is strictly positive and $p_M$ is finite so that the ratio $\frac{P_M}{p_m^2}$ is bounded.
Since this ratio essentially determines the exponential rate of convergence of the estimation errors to zero, it is of interest to know bounds of $p_m$ and $p_M$ in relation to the "amount" of observability. Such bounds are derived in Appendix \ref{bounds}, with a lower bound of $p_M$ calculated from an expression derived in \cite{PRV2001}.
The central issue of boundedness and good conditioning of $P(t)$ brings us to recall classical, and also point out less known, results concerning the CRE.

\subsection{Properties of the Continuous Riccati Equation}
The first results concerns the existence of the solutions to the CRE for $t \in [0,+\infty)$.
\begin{lemma} \label{existence}
If $P(0)$ is p.d. and $Q(t)$ and $V(t)$ are p.s.d, then $P(t)$ is p.d. and well defined on $[0,+\infty)$.
\end{lemma}
See the proof in Appendix \ref{proof-existence}.

Now, to ensure boundedness and good-conditioning of the solution $P(t)$ to the CRE one has to impose other conditions upon the terms entering the equation. Sufficient conditions are pointed out in the next lemma.

\begin{lemma} \label{sufficient_conditions}
Define:
\begin{equation} \label{Wv}
W_V(t,t+\delta):=\frac{1}{\delta} \int_t^{t+\delta}\Phi(t,s)V(s)\Phi^{\top}(t,s)ds
\end{equation}
and
\begin{equation} \label{Wq}
W_Q(t,t+\delta):=\frac{1}{\delta} \int_t^{t+\delta}\Phi^{\top}(s,t)C^{\top}(s)Q(s)C(s)\Phi(s,t)ds
\end{equation}
If there exist (strictly) positive numbers $\delta$, $\mu_v$, and  $\mu_q$ such that $W_V(t,t+\delta) \geq \mu_v I_d$ and $W_Q(t,t+\delta) \geq \mu_q I_d$, $\forall t$, then the solution $P(t)$ to the CRE \eqref{dP} is bounded and well-conditioned in the sense that $0<p_m \leq p_M<\infty$.
\end{lemma}

A proof of this result is given in \cite{PRV2001} where lower and upper bounds of $P(t)$ are also derived. The matrix $Q(t)$ is in fact assumed p.d. because the inverse of $Q$ is (for technical convenience) used in the proof. However, it is simple to verify that the proposed bounds for $P(t)$ 
do not depend on the smallest eigenvalue of $Q(t)$, so that these bounds are also valid when $Q(t)$ is only p.s.d.

From now on, and by analogy with the previously defined observability Grammian $W$ of System \eqref{linear_system}, $W_Q$ is called {\em Riccati observability Grammian}. It coincides with $W$ when $Q(t)=I_d$. Note that if $Q(t)\geq \epsilon I_d>0$  and the observability Grammian $W$ satisfies the positivity condition \eqref{grammian}, then the Riccati observability Grammian $W_Q$ satisfies a similar condition. This is just a consequence of that $W_Q(t,t+\delta) \geq \lambda_{min}(Q(t))W(t,t+\delta)$.

The above lemma calls for the following (well known) corollaries whose proofs are simple and omitted here for the sake of conciseness.

\begin{corollary} \label{corollary1}
~

\begin{itemize}
\item If $A$ and $B$ are constant and such that the pair $(A,B)$ is (Kalman) controllable, and if $V(t)=B\bar{V}(t)B^{\top}$ with $\bar{V}(t) \geq \epsilon I_d>0$, $\forall t$, then the condition upon $W_V$ is satisfied. These conditions are themselves satisfied in the particular case where $B=I_d$ and $V(t)\geq \epsilon I_d>0$.
\item If $A$ and $C$ are constant and such that the pair $(A,C)$ is (Kalman) observable, and if $Q(t)\geq \epsilon I_d>0$, $\forall t$, then the condition upon $W_Q$ is satisfied.
\end{itemize}
\end{corollary}
It is however useful to keep in mind that the above conditions for the boundedness and good-conditioning of $P(t)$ are {\em only sufficient}. For instance, when $A$ is constant and Hurwitz stable, it is also sufficient to take $Q(t)=0$, $\forall t$ (so that the Riccati observability Grammian is identically equal to zero) and $V$ constant. Indeed, it is simple to show that $P(t)$ then converges to the p.d. solution to the Lyapunov equation $AP+PA^{\top}+V=0$.

A second corollary, that will be used further for the design of a position observer based on direction measurements, is as follows
\begin{corollary} \label{corollary2}
Consider the projection matrix operator $\Pi_{y(t)}:=I_d-y(t)y^{\top}(t)$ with $y(t) \in \rr^n$ and such that $|y(t)|=1$ (i.e. $y(t) \in S^{n-1}$). If $A(t)$ is the null matrix, $C(t)=\Pi_{y(t)}$ and  $V(t)\geq \epsilon_v I_d>0$ or $V(t)=k_v \Pi_{y(t)}$, with $\epsilon_v$ and $k_v$ denoting positive numbers, and if the following {\em persistent excitation (p.e.) condition} is satisfied
\begin{equation} \label{persistent}
\forall t: \frac{1}{\delta}\int_t^{t+\delta}\Pi_{y(s)}ds \geq \epsilon I_d >0~~,~\mbox{for some}~\delta>0
\end{equation}
then the conditions on $W_V$ and $W_Q$ are also satisfied (and the solution $P(t)$ to the CRE is thus bounded and well conditioned).
\end{corollary}
If one chooses $Q(t)=k_q I_d$ and $V(t)=k_v \Pi_{y(t)}$ then the CRE writes as
\[
\dot{P}=-k_qP\Pi_{y(t)}P+k_v\Pi_{y(t)}~;~P(0):p.d.
\]
and $P(t)$ converges to $(k_v/k_q)^{0.5}I_d$. Note that this latter matrix is a solution to the CRE even when the condition \eqref{persistent} of persistent excitation is not satisfied.

A technical extension of this corollary, also used further for the same estimation problem, but in the case where the velocity measurement is biased, follows
\begin{lemma} \label{extension}
If \begin{enumerate}
\item $C(t)=\Pi_{y(t)} \bar{C}$ with $\bar{C}$ a constant matrix,
\item $A$ is constant and such that the pair $(A,\bar{C})$ is Kalman observable,
\item all eigenvalues of $A$ are real, i.e. $det(\lambda I_d-A)=0 ~\Rightarrow~\lambda \in \rr$,
\item the p.e. condition \eqref{persistent} is satisfied,
\end{enumerate}
then the Riccati observability Grammian $W_Q(t,t+\bar{\delta})$ is positive for some $\bar{\delta}>0$ and $\forall t \geq 0$.
\end{lemma}
See the proof in Appendix \ref{proof-extension}.
Let us just remark that the requirement of eigenvalues of $A$ being all real is not fortuitous. Indeed, it is not difficult to find out examples for which the non-satisfaction of this condition forbids the Riccati observability Grammian from being positive. Such an example is
\[
A=\left[ \begin{array}{cc} 0 & 1 \\ -1 & 0 \end{array} \right]~~,~\bar{C}=I_d
\]
The pair $(A,\bar{C})$ is clearly observable, and $det(\lambda I_d-A)=\lambda^2+1$ so that the eigenvalues of $A$ are the imaginary numbers $\pm j$. Choose $y(t)=[\cos(t),-\sin(t)]^{\top}$ so that
\[
\begin{array}{lll}
\int_t^{t+2\pi}\Pi_{y(s)}ds & = & \int_t^{t+2\pi}\left[ \begin{array}{cc} \sin(s)^2 & \sin(2s)/2 \\ \sin(2s)/2 & \cos(s)^2 \end{array} \right]ds \\
~ & = &\pi I_d
\end{array}
\]
This establishes that the p.e. property \eqref{persistent} is satisfied.
However, using the fact that $exp(At)=\cos(t)I_d+\sin(t)A$ and setting $b=[1,0]^{\top} \in S^1$, one verifies that
\[
\Pi_{y(s)}\bar{C}exp(A(s))b=\Pi_{y(s)}y(s)=0~~,~\forall s
\]
This proves that the Riccati observability Grammian is never invertible in this case.

\section{Observers for position estimation from direction measurements}\label{sec:direction}
The problem consists in estimating the position $x$ of a body (or object) with respect to (w.r.t.) a fixed frame given its velocity $u$ and the measurement of its direction $x/|x|$, knowing that the measured velocity may be biased by an initially unknown constant vector $a$. In practice, $x$ will be a two-dimensional vector of coordinates (in the 2D, or planar, case) or a three-dimensional vector of coordinates (in the 3D, or spatial, case). For the sake of generality, we assume here that $x \in \rr^n$, with $n\geq 2$.
The corresponding modelling equations are
\begin{equation} \label{model_direction}
\begin{array}{l}
\dot{x}=u+a \\
\dot{a}=0 \\
0=\Pi_{y(t)}x
\end{array}
\end{equation}
with $y(t):=x(t)/|x(t)|$.
Let us distinguish two cases, depending on whether the velocity measurement is unbiased, i.e. $a=0$, or is
biased by an unknown constant vector $a$.

\subsection{Unbiased velocity measurements}
In this case the modelling equations can also be written as
\begin{equation} \label{model_direction2}
\begin{array}{l}
\dot{x}=Ax+u \\
0=C(t)x
\end{array}
\end{equation}
with $A=0_{n \times n}$ --the $(n\times n)$-dimensional null matrix-- and $C(t)=\Pi_{y(t)}$. This system can be associated with the following Riccati-like observer
\begin{equation} \label{observer_direction_unbiased}
\dot{\hat{x}}=A\hat{x}+u+K(t)(0-C(t)\hat{x})
\end{equation}
with the observer gain $K(t)$ calculated as in relation \eqref{riccati-gain} from the solution to the CRE \eqref{dP}. 
The resulting observer writes as
\begin{equation} \label{observer_direction_unbiased2}
\dot{\hat{x}}=u-K(t)\Pi_{y(t)}\hat{x}
\end{equation}
with $K(t)=k(t)P(t)\Pi_{y(t)}Q(t)$ and $P(t)$ the solution to the CRE
\[
\dot{P}=-P\Pi_{y(t)}Q(t)\Pi_{y(t)}P+V(t)
\]
Since $\dot{\tilde{x}}=(A-K(t)C(t))\tilde{x}$ the Lyapunov analysis of section \ref{riccati-observers} applies, and global exponential stability of $\tilde{x}=0$ is obtained provided that $P(t)$ is bounded and well-conditioned. From Corollary \ref{corollary2} such is the case if $V(t)$ is chosen positive (and larger than $\epsilon_v I_d$) or equal to $k_v \Pi_{y(t)}$, and --most importantly-- if the p.e. condition \eqref{persistent} is satisfied. A loose interpretation of this condition is that the body must keep moving and not always in the direction of the source point. Note that, in the case where $V(t)=k_v \Pi_{y(t)}$ and $Q(t)=k_q I_d$, choosing the constant solution $P=(k_v/k_q)^{0.5}I_d$ simplifies the observer equation to $\dot{\hat{x}}=u-k(t)\sqrt{k_qk_v}\Pi_{y(t)}\hat{x}$, so that one recovers the solution proposed in \cite{LebHamMahSam15}. 

\subsection{Biased velocity measurements}
In this more difficult case the velocity bias $a$ has to be estimated as well. The modelling equations \eqref{model_direction} can be written in the state form as
\begin{equation} \label{model_direction3}
\begin{array}{l}
\dot{X}=AX+\bar{u} \\
0=C(t)X
\end{array}
\end{equation}
with $X:=[x^{\top},a^{\top}]^{\top}$ the $2n$-dimensional extended state vector, $\bar{u}:=[u^{\top},0_{1\times n}]^{\top}$, and
\[
A=\left[ \begin{array}{cc} 0_{n\times n} & I_d \\ 0_{n\times n} & 0_{n\times n} \end{array} \right],~C=\Pi_{y(t)}\bar{C},~\bar{C}=\left[ \begin{array}{cc} I_d & 0_{n\times n} \end{array} \right]
\]
Note that the pair $(A,\bar{C})$ is Kalman-observable since $[\bar{C}^{\top},(\bar{C}A)^{\top}]$ is equal to the $(2n\times 2n)$-dimensional identity matrix. Consider now the CRE
\begin{equation} \label{CRE-bias}
\dot{P}=AP+PA^{\top}-PC^{\top}(t)Q(t)C(t)P+V(t)
\end{equation}
with $P(0)$: a p.d. matrix.
Provided that the p.e. condition \eqref{persistent} is satisfied, the solution $P(t)$ to this CRE is bounded and well-conditioned, by application of Lemma \ref{extension} after noticing that all eigenvalues of $A$ are equal to zero (and thus real). Consider now the following Riccati observer
\begin{equation} \label{observer_direction_biased}
\dot{\hat{X}}=A\hat{X}+\bar{u}+K(t)(0-C(t)\hat{X})
\end{equation}
with $\hat{X}:=[\hat{x}^{\top},\hat{a}^{\top}]^{\top}$ and $K(t)=k(t)P(t)C^{\top}(t)Q(t)$ ($k(t)\geq 0.5$). This observer can also be written as
\begin{equation} \label{observer_direction_single}
\left\{ \begin{array}{l}
\dot{\hat{x}}=u+\hat{a}-k(t)P_{11}(t)\Pi_{y(t)}Q(t)\hat{x} \\
\dot{\hat{a}}=-k(t)P_{21}(t)\Pi_{y(t)}Q(t)\hat{x}
\end{array} \right.
\end{equation}
with $P_{ij}$ ($i,j \in \{1,2\}$) denoting the block components of $P$ with adequate dimensions. Since the estimation error satisfies the equation $\dot{\tilde{X}}=(A-K(t)C(t))\tilde{X}$, the Lyapunov analysis of section \ref{riccati-observers} applies and global exponential stability of $\tilde{X}=0$ is obtained provided that the p.e. condition \eqref{persistent} is satisfied.

\subsection{Extension to multiple direction measurements}
We consider now the problem of estimating a vector $x$ from $l$ measurements $y_i=\frac{x-z_i}{|x-z_i|}$, $i\in\{1,\ldots,l\}$. In the case where $x$ represents the position of a body w.r.t. an inertial frame, and $z_i$ is the known vector of coordinates of a fixed source point, then $y_i$ is the unit vector measuring the direction between the object and this point.\\
Setting $X:=[x^{\top},a^{\top}]^{\top}$,$\bar{u}:=[u^{\top},0_{1\times n}]^{\top}$, and $y=[(\Pi_{y_1(t)}z_1)^{\top},\ldots,(\Pi_{y_l(t)}z_l)^{\top}]^{\top}$, one obtains the system 
\begin{equation} \label{model_biased_multi}
\begin{array}{l}
\dot{X}=AX+\bar{u} \\
y=C(t)X
\end{array}
\end{equation}
with
\[
\begin{array}{lll}
A &=&\left[ \begin{array}{cc} 0_{n\times n} & I_d \\ 0_{n\times n} & 0_{n\times n} \end{array} \right] \\
C(t)&=&\left[ \begin{array}{cccc}
\Pi_{y_1(t)} & 0_{n\times n} & \ldots & 0_{n\times n} \\
0_{n\times n} & \Pi_{y_2(t)} & \ldots & 0_{n\times n} \\
\vdots & \vdots & ~ & \vdots \\
0_{n\times n} & 0_{n\times n} & \ldots & \Pi_{y_l(t)}
\end{array} \right]\bar{C} \\
\bar{C}&=&\left[ \begin{array}{cc} I_d & 0_{n\times n} \\ \vdots \\I_d & 0_{n\times n} \end{array} \right]~\mbox{with}~ dim(\bar{C})=ln \times 2n
\end{array}
\]
It is simple to verify that the pair $(A,\bar{C})$ is Kalman-observable.
Consider now the CRE
\[
\dot{P}=AP+PA^{\top}-PC^{\top}(t)Q(t)C(t)P+V(t)
\]
with $P(0)$: a p.d. matrix and
\[
Q(t)=\left[ \begin{array}{cccc}
Q_{11}(t) & 0_{n\times n} & \ldots & 0_{n\times n} \\
0_{n\times n} & Q_{22}(t) & \ldots & 0_{n\times n} \\
\vdots & \vdots & ~ & \vdots \\
0_{n\times n} & 0_{n\times n} & \ldots & Q_{ll}(t)
\end{array} \right]
\]
The solution $P(t)$ to this equation is bounded and well conditioned provided that the corresponding Riccati observation Grammian $W_Q$ is positive. Using the fact that
\[
C^{\top}(t)Q(t)C(t)=\left[ \begin{array}{cc}
\Delta(t) & 0_{n\times n} \\
0_{n\times n} & 0_{n\times n}
\end{array} \right]
\]
with $\Delta(t):=\sum_{i=1}^l \Pi_{y_i(t)}Q_{ii}(t) \Pi_{y_i(t)}$, one verifies that this condition is satisfied if, for some $\delta >0$ and for all $t>0$,
$\frac{1}{\delta}\int_t^{t+\delta}\bar{\Delta}(s)ds$, with $\bar{\Delta}(t):=\sum_{i=1}^l \Pi_{y_i(t)}$, is greater than an arbitrarily small s.p.d. matrix. This p.e. condition clearly points out the interest of using multiple direction measurements in order to weaken, or even remove, conditions upon $x$ and its time-variations. For instance, in the 3D case ($n=3$), if $l \geq 2$ then this p.e. condition is satisfied provided that the body is periodically not aligned with all the source points. If three or more source points are not aligned, then this condition is automatically satisfied independently of $x$ and its time-variations.\\
A Riccati observer associated with this system is 
\begin{equation} \label{observer_direction_biased_multi}
\dot{\hat{X}}=A\hat{X}+\bar{u}+K(t)(y-C(t)\hat{X})
\end{equation}
with $\hat{X}:=[\hat{x}^{\top},\hat{a}^{\top}]^{\top}$ and $K(t)=k(t)P(t)C^{\top}(t)Q(t)$ ($k(t)\geq 0.5$).
One easily verifies that this observer can also be written as
\begin{equation} \label{observer_direction_multiple}
\left\{ \begin{array}{l}
\dot{\hat{x}}=u+\hat{a}-k(t)P_{11}(t)(\sum_{i=1}^l \Pi_{y_i(t)}Q_{ii}(t)(\hat{x}-z_i)) \\
\dot{\hat{a}}=-k(t)P_{21}(t)(\sum_{i=1}^l \Pi_{y_i(t)}Q_{ii}(t)(\hat{x}-z_i))
\end{array} \right.
\end{equation}
From what precedes this observer globally exponentially stabilizes $\tilde{X}=0$ if the previously evoked p.e. condition is satisfied.\\~\\
{\em Remarks}:\\
$\bullet$ In the 3D-case, if $l \geq 2$ and the matrix $\bar{\Delta}(t)$ is positive, and if the body moves with a constant unknown velocity, the above observer provides also an estimation of this velocity. To this aim it suffices to set $u=0$ in the algorithm. The term $\hat{a}$ is then an estimate of the body velocity that is equal to $a$ in this case.\\
$\bullet$ In the unbiased case where $a=0$ and the body velocity $u$ is measured, the calculation of $\hat{a}$ is superfluous and the above observer reduces to
\[
\dot{\hat{x}}=u-k(t)P(t)(\sum_{i=1}^l \Pi_{y_i(t)}Q_{ii}(t)(\hat{x}-z_i))
\]
with $P(t)$ the solution to the CRE
\[
\dot{P}=-P\Delta(t)P+V(t)
\]
A particular solution to this latter equation, obtained by choosing $V(t)=\Delta(t)$, is $P=I_d$.

\section{Observers for position estimation from range measurements}\label{sec:range}
The problem consists in estimating the position $x$ of a body w.r.t. a fixed frame given its velocity $u$ and the measurement of the distance (or range) $|x|$, knowing that the measured velocity may be biased by an initially unknown constant vector $a$. Again, for the sake of generality, we assume that $x \in \rr^n$, with $n\geq 2$. For the sake of simplifying forthcoming relations, and also to prevent the designed observer from being singular when $|x|=0$, it is useful to formally set the system's output $y$ equal to $0.5 |x|^2$ rather than $|x|$.
The corresponding modelling equations are
\begin{equation} \label{model_range}
\begin{array}{l}
\dot{x}=u+a \\
\dot{a}=0 \\
y=0.5 |x|^2
\end{array}
\end{equation}
Let us again distinguish two cases, depending on whether the velocity measurement is unbiased, i.e. $a=0$, or is
biased by an unknown constant vector $a$.

\subsection{Unbiased velocity measurements}
In this case the modelling equations can be linearised by defining the $(n+1)$-dimensional extended state vector $X:=[x^{\top},y]^{\top}$. Indeed, forming the time-derivative of $X$ yields the LTV system
\begin{equation} \label{model_range2}
\begin{array}{l}
\dot{X}=A(t)X+\bar{u} \\
y=CX
\end{array}
\end{equation}
with $\bar{u}:=[u^{\top},0]^{\top}$ and
\[
A(t)=\left[ \begin{array}{cc} 0_{n\times n} & 0_{n\times 1} \\ u^{\top}(t) & 0 \end{array} \right]~~,~C=[0_{1\times n}~ 1]
\]
A Riccati observer associated with this system is
\begin{equation} \label{observer_range_unbiased}
\begin{array}{l}
\dot{\hat{X}}=A\hat{X}+\bar{u}+K(t)(y-C\hat{X}) \\
K(t)=k(t)P(t)C^{\top}q(t) \\
\dot{P}=A(t)P+PA^{\top}(t)-PC^{\top}q(t)CP+V(t)
\end{array}
\end{equation}
with $P(0)$ a p.d. matrix, $k(t)\geq 0.5$, $q(t)\geq \epsilon > 0$, and $V(t)\geq \epsilon I_d>0$. Setting $\hat{X}=[\hat{x}^{\top},\hat{y}]^{\top}$ this observer can equivalently be written as
\[
\left\{ \begin{array}{l}
\dot{\hat{x}}=u(t)+k(t)q(t)P_{21}(t)(y(t)-\hat{y}(t)) \\
\dot{\hat{y}}=u^{\top}(t)\hat{x}+k(t)q(t)p_{22}(t)(y(t)-\hat{y}(t))
\end{array} \right.
\]
This observer globally exponentially stabilizes the estimation error $\tilde{X}:=X-\hat{X}$ at zero if $P(t)$ is bounded and well-conditioned. From what precedes such is the case if there exists $\delta>0$ and $\mu_q>0$ such that the Riccati observability Grammian satisfies $W_Q(t,t+\delta)\geq \mu_q I_d$, $\forall t \geq 0$. We claim that this latter condition is itself satisfied when the p.e. condition upon $u(t)$ specified in the next lemma is satisfied.

\begin{lemma} \label{excitation_range_unbiased}
If $u(t)$ satisfies the p.e. condition
\begin{equation} \label{peu}
\forall t\geq 0~:~\frac{1}{\delta}\int_{t}^{t+\delta}u(s)u^{\top}(s) ds \geq \mu I_d >0
\end{equation}
for some $\delta>0$ and $\mu>0$, then the Riccati observer \eqref{observer_range_unbiased} globally exponentially stabilizes $\tilde{X}=0$.
\end{lemma}
See the proof in Appendix \ref{proof_excitation_range_unbiased}.
~\\
\noindent {\em Remark:} Another observer yielding the asymptotic stability of $\tilde{X}=0$ under the p.e. condition upon $u(t)$ and when $u(t)$ is uniformly continuous is
\[
\left\{ \begin{array}{l}
\dot{\hat{x}}=u(t)+k_1(y(t)-\hat{y}(t)) \\
\dot{\hat{y}}=u^{\top}(t)\hat{x}+k_2(t)(y(t)-\hat{y}(t))
\end{array} \right.
\]
with $k_1>0$ and $0\leq \underline{k_2} \leq k_2(t) \leq \overline{k_2} < \infty$.
This can be proved by considering the positive function $\V(t):=|\tilde{x}(t)|^2/k_1+\tilde{y}(t)^2$ whose time-derivative satisfies $\dot{\V}(t)=-2k_2(t)\tilde{y}(t)^2$ ($\leq 0$). One deduces that $\tilde{y}(t)$ converges to zero and that $|\tilde{x}(t)|$ is bounded and converges to some finite limit $|\tilde{x}_{\infty}|$. Then, by application of (extended) Barbalat's Lemma one deduces that the time-derivative of $\tilde{y}(t)$ converges to zero so that $u^{\top}(t)\tilde{x}(t)$ also converges to zero. Therefore $\int_t^{t+\delta} |u^{\top}(s)\tilde{x}(s)|^2ds$ converges to zero when $t$ tends to infinity. The convergence of $\tilde{y}(t)$ to zero also implies that the time-derivative of $\tilde{x}(t)$ converges to zero. From there one finishes the proof by showing that the satisfaction of the p.e. condition is not compatible with $|\tilde{x}_{\infty}| \neq 0$.

An advantage of this second observer over a Riccati observer is that it involves less calculations. However, the proof of convergence of this observer, as sketched above, does not establish that the rate of convergence is exponential. This limitation epitomizes an important feature that goes with Riccati observer designs, namely the knowledge of an explicit Lyapunov function that allows for a more complete analysis of stability and convergence.

\subsection{Biased velocity measurements}
In this more difficult case the modelling equations can be linearised by defining the $(2n+3)$-dimensional extended state vector $X:=[x^{\top},a^{\top},y,a^{\top}x,|a|^2]^{\top}$. Indeed, forming the time-derivative of $X$ yields the LTV system
\begin{equation} \label{model_range3}
\begin{array}{ll}
\dot{X}=A(t)X+\bar{u} \\
y=CX
\end{array}
\end{equation}
with $\bar{u}:=[u^{\top},0_{1\times(n+3)}]^{\top}$ and
\[
\begin{array}{c}
A(t) = \left[ \begin{array}{ccccc} 0_{n\times n} & I_{n\times n} & 0_{n\times 1} & 0_{n\times 1} & 0_{n\times 1}\\
0_{n\times n} & 0_{n\times n} & 0_{n\times 1} & 0_{n\times 1} & 0_{n\times 1} \\
u^{\top}(t) & 0_{1\times n} & 0 & 1 & 0 \\
0_{1 \times n} & u^{\top}(t) & 0 & 0 & 1 \\
0_{1\times n} & 0_{1\times n} & 0 & 0 & 0
\end{array} \right]
\\
C = [0_{1\times n}~0_{1\times n}1~0~0]
\end{array}
\]
A Riccati observer associated with this sytem is given by \eqref{observer_range_unbiased}, and global exponential stability of this observer follows if the system's observability Grammian satisfies \eqref{grammian}. We claim that this latter condition is itself satisfied when the p.e. condition upon $\dot{u}(t)$ specified in the next lemma is satisfied.

\begin{lemma} \label{excitation_range_biased}
If $u(t)$ is twice differentiable with bounded first and second derivatives and if $\dot{u}(t)$ satisfies the p.e. condition
\begin{equation} \label{pedu}
\forall t\geq 0~:~\frac{1}{\delta}\int_{t}^{t+\delta}\dot{u}(s)\dot{u}^{\top}(s) ds \geq \mu I_d >0
\end{equation}
for some $\delta>0$ and $\mu>0$, then the Riccati observer globally exponentially stabilizes $\tilde{X}=0$.
\end{lemma}
See the proof in Appendix \ref{proof_excitation_range_biased}.

\subsection{Extension to multiple range measurements}
We consider now the problem of estimating a vector $x$ from $l$ measurements $y_i=0.5|x-z_i|^2$, $i\in\{1,\ldots,l\}$. In the case where $x$ represents the 3D position of a body w.r.t. an inertial frame, and $z_i$ is the known vector of coordinates of a fixed source point, then $y_i$ is half the squared distance between the body and this point.
\subsubsection{Unbiased velocity measurements ($a=0$)}
Define the $(l \times 1)$-dimensional constant vector $\xi:=[1,\ldots,1]^{\top}$. Define the {\em weighted} output variable $y_0:=\sum_{i=1}^l\alpha_i (y_i-0.5|z_i|^2)$, with $\alpha=[\alpha_1,\ldots,\alpha_l]^{\top}$ denoting a $l$-dimensional vector of real numbers such that $\sum_{i=1}^l\alpha_i=1$. Since $\dot{x}=u$ and
$y_0=0.5|x|^2-\sum_{i=1}^l\alpha_i z_i^{\top}x$, one has $\dot{y}_0=(x^{\top}-\sum_{i=1}^l\alpha_i z_i^{\top})u$.
Define also the $l$-dimensional output vector $y:=[y_0,(y_1-y_0-0.5|z_1|^2),\ldots,(y_l-y_0-0.5|z_l|^2)]^{\top}$ and the augmented state $X:=[x^{\top},y_0]^{\top}$. Since $(y_j-y_0-0.5|z_j|^2)=\sum_{i=1}^l\alpha_i z_i^{\top}x-\alpha_jz_j^{\top}x$ one has $y=CX$ with
\[
\begin{array}{ll}
C:=\left[ \begin{array}{cc}
0_{1 \times n} & 1 \\
D(\alpha) Z^{\top} & 0_{l \times 1}
\end{array} \right] & ~
\\
D(\alpha):=\xi \alpha^{\top}-I_{l \times l} &~: (l\times l)\mbox{-dimensional matrix}
\\
\xi:=[1,\ldots,1]^{\top} &~: (l\times 1)\mbox{-dimensional vector}
\\
Z:=[z_1 \ldots z_l] &~:(n\times l)\mbox{-dimensional matrix}
\end{array}
\]
Note that the rank of the matrix $D(\alpha)$ is equal to $(l-1)$.
From the previous definitions one obtains the linear system
\begin{equation} \label{system-multiple-range}
\begin{array}{lll}
\dot{X} & = & A(t)X+\bar{u}\\
y & = & CX
\end{array}
\end{equation}
with
\[
A(t):=\left[ \begin{array}{cc}
0_{n \times n} & 0_{n \times 1} \\
u(t)^{\top} & 0
\end{array} \right]~,
~\bar{u}:=\left[ \begin{array}{c}
I_{n \times n} \\ -\sum_{i=1}^l\alpha_i z_i^{\top}
\end{array} \right] u
\]
A Riccati observer associated with this system is of the form \eqref{observer_range_unbiased}, except for the positive scalar $q(t)$ involved in the CRE that is now replaced by a $((l+1) \times (l+1))$-dimensional p.d. matrix $Q(t)$.

\begin{lemma} \label{excitation_multiple_range_unbiased}
If $u(t)$ and the vectors $z_i$ ($i=1,\ldots,l$) satisfy the p.e. condition
\begin{equation} \label{peu}
\forall t\geq 0~:~ZD^{\top}(\alpha)D(\alpha)Z^{\top}+\frac{1}{\delta}\int_{t}^{t+\delta}u(s)u^{\top}(s) ds \geq \mu I_d
\end{equation}
for some $\delta>0$ and $\mu>0$, then the above-mentioned Riccati observer globally exponentially stabilizes $\tilde{X}=0$.
\end{lemma}
The proof of this lemma is a straightforward adaptation of the proof of Lemma \ref{excitation_range_unbiased} given in Appendix \ref{proof_excitation_range_unbiased}, after observing that the matrix $M(t)$ involved in the proof is now
\[
M(t)=\left[ \begin{array}{cc}
0_{1 \times n} & 1 \\
D(\alpha)Z^{\top} & 0_{l \times 1} \\
u^{\top}(t) & 0
\end{array} \right]~.
\]
We remark that the p.e. condition is automatically satisfied, independently of $u(t)$, when $l \geq n+1$ and $rank(D(\alpha)Z^{\top})=n$, i.e. when $n$ vectors among the $l$ vectors $z_j-\sum_{i=1}^l\alpha_i z_i$ ($j=1,\ldots,l$) are independent. For instance, in the 3D case  (resp. 2D case) it is satisfied when the number of source points is equal to, or greater than, four (resp. three) and all source points are not coplanar (resp. aligned). This result is coherent with the minimum number of source points needed to geometrically determine the position of a motionless body {\em with no ambiguity} from a single set of multiple range measurements. Using more source points provides redundancy that can be used to accelerate the rate of convergence and/or reduce the asymptotic variance of $\tilde{X}$ when the range measurements are corrupted by noise. More generally, Riccati observers performance depends on the tuning of the parameters involved in the CRE, namely $k(t)$, $Q(t)$, and $V(t)$. In this respect the choice of parameters associated with the stochastic optimal Kalman filter can provide useful leads in complementation with the dependence, pointed out earlier, between the exponential convergence rate of the observer and the amount of persistent excitation. This tuning issue is important for practical purposes and deserves to be studied in its own right. However, it is out of the present paper's scope and is thus not pursued further here.

\subsubsection{Biased velocity measurements ($a$ is {\em a priori} unknown)}
Define
\begin{itemize}
\item $X:=[x^{\top},a^{\top},y_0,a^{\top}x,|a|^2]^{\top}$;
\item $y:=[y_0,(y_1-y_0-0.5|z_1|^2),\ldots,(y_l-y_0-0.5|z_l|^2)]^{\top}$, i.e. the same output vector as in the unbiased case;
\item  $\bar{u}:=[u^{\top},0_{1 \times n},-u^{\top}(\sum_{i=1}^l\alpha_i z_i),0,0]^{\top}$.
\end{itemize}
Forming the time-derivative of $X$ yields a linear system alike \eqref{system-multiple-range}
with the state matrix
\[
A(t) = \left[ \begin{array}{ccccc} 0_{n\times n} & I_{n\times n} & 0_{n\times 1} & 0_{n\times 1} & 0_{n\times 1}\\
0_{n\times n} & 0_{n\times n} & 0_{n\times 1} & 0_{n\times 1} & 0_{n\times 1} \\
u^{\top}(t) & -\sum_{i=1}^l\alpha_i z_i^{\top} & 0 & 1 & 0 \\
0_{1\times n} & u^{\top}(t) & 0 & 0 & 1 \\
0_{1\times n} & 0_{1\times n} & 0 & 0 & 0
\end{array} \right]
\]
and the output matrix
\[
C:=\left[ \begin{array}{ccccc}
0_{1 \times n} & 0_{1 \times n} & 1 & 0 & 0 \\
D(\alpha)Z^{\top} & 0_{1 \times n} & 0_{l \times 1} & 0_{l \times 1} & 0_{l \times 1}
\end{array} \right]
\]
A Riccati observer associated with this system is thus again given by \eqref{observer_range_unbiased}, with the p.d. matrix $Q(t)$ involved in the CRE chosen as in the unbiased case with multiple range measurements.

\begin{lemma} \label{excitation_multiple_range_biased}
If $u(t)$ is twice differentiable with bounded first and second derivatives, and if $\dot{u}(t)$ and the vectors $z_i$ ($i=1,\ldots,l$) satisfy the p.e. condition
\begin{equation} \label{peduz}
\forall t\geq 0~:~ZD^{\top}(\alpha)D(\alpha)Z^{\top}+\frac{1}{\delta}\int_{t}^{t+\delta}\dot{u}(s)\dot{u}^{\top}(s) ds \geq \mu I_d 
\end{equation}
for some $\delta>0$ and $\mu>0$, then the Riccati observer globally exponentially stabilizes $\tilde{X}=0$.
\end{lemma}
The proof of this lemma is a simple adaptation of the proof of Lemma \ref{excitation_range_biased} given in Appendix \ref{proof_excitation_range_biased}, with the matrix $M(t)$ involved in the proof chosen as follows
\[
M(t)=\left[ \begin{array}{c}N_0 \\ N_1(t) \\ \bar{N}_2(t) \end{array} \right]
\]
with $N_0=C$, $N_1(t)=CA(t)$, and $\bar{N}_2(t)$ equal to the first line of $N_2(t)=N_1(t)A(t)+\dot{N}_1(t)$.
According to this lemma one finds again that the positivity of the matrix $ZD^{\top}(\alpha)D(\alpha)Z^{\top}$, which is generically ensured when the number of source points is greater than three in the 3D-case (resp. two in the 2D-case), is sufficient to yield the exponential stabilization of $\tilde{X}=0$ independently of the input $u(t)$. Nevertheless, the knowledge of the time-derivative of $x$, via the estimation of the bias $a$, is still required.\\~\\
{\em Remark}: When the rank of the matrix $D(\alpha)Z^{\top}$ is equal to $n$, i.e. when using at least four non-coplanar source points in the 3D-case and at least three non-aligned source points in the 2D-case, and when the body velocity is constant but unknown {\em a priori}, the observer provides also an estimate of this velocity. To this aim it suffices to set $u=0$ in the algorithm. The term $\hat{a}$ is then an estimate of the body velocity $a$.

\subsubsection{Biased range measurements} \label{biased-range}
A practical reason for considering the case of range measurements corrupted by an additive constant bias $b$ stems from that several range sensors measure times of flight via the use of clocks that are not necessarily exactly synchronized. For instance, in the case of GNSS (global navigation satellite systems) the clocks of the satellites are typically desynchronized from the receiver's clock by a small value $\Delta t$ that produces a range measurement bias equal to $c\Delta t$, with $c$ denoting the speed of light. This leads to estimate $b$ together with the body position $x$.\\
Let us thus assume that the measured distance is $\bar{y}_i=|x-z_i|+b$ ($i\in \{1,\ldots,l\}$) with $b$ an unknown real number. By analogy with the unbiased case let us set $y_i:=0.5 \bar{y}_i^2$ for $i\in \{1,\ldots,l\}$ and $y_0:=\sum_{i=1}^l\alpha_i (y_i-0.5|z_i|^2)$, with $\alpha=[\alpha_1,\ldots,\alpha_l]^{\top}$ denoting a $l$-dimensional vector of real numbers such that $\sum_{i=1}^l\alpha_i=1$. The (measured) output vector is still $y:=[y_0,(y_1-y_0-0.5|z_1|^2),\ldots,(y_l-y_0-0.5|z_l|^2)]^{\top}$. Define $\bar{y}:=[\bar{y}_1,\ldots,\bar{y}_l]^{\top}$, the augmented state vector as $X:=[x^{\top},y_0-b(\alpha^{\top}\bar{y}),b]^{\top} \in \rr^{n+2}$ and the input vector $\bar{u}:=[u^{\top},-\sum_{i=1}^l\alpha_i(z_i^{\top}u),0]^{\top}$. One obtains the following LTV system: $\dot{X}=A(t)X+\bar{u}$, $y=C(t)X$, with
\vspace{-0.3cm}
\[
A(t) = \left[ \begin{array}{ccc} 0_{n\times n}  & 0_{n\times 1} & 0_{n\times 1}\\
u^{\top}(t) & 0 & 0 \\
0_{1\times n} & 0 & 0
\end{array} \right]
\]
and
\[
C=\left[ \begin{array}{ccc}
0_{1 \times n} & 1 & \alpha^{\top}\bar{y} \\
D(\alpha)Z^{\top} & 0_{l \times 1} & -D(\alpha)\bar{y}
\end{array} \right]
\]
A Riccati observer associated with this system is of the form \eqref{observer_direction_biased_multi}, with the matrix $Q(t)$ chosen larger than an arbitrarily small p.d. matrix.
\begin{lemma} \label{proposition}
~

\begin{enumerate}
\item {\em Motionless body} (static case):\\
Let $d_i=|x-z_i|$ ($i=1,\ldots,l$) denote the distance between the $ith$ source point to the body, and $r_i:=(x-z_i)/|x-z_i|$ ($i \in \{1,\ldots,l\}$) denote the unit vector charactering the direction from the $ith$ source point to the body. Assume that at least $(n+2)$ source points are used and that the rank of $D(\alpha)Z^{\top}$ is equal to $n$ (a condition satisfied as soon as four non-coplanar source points (in the 3D-case), or three non-aligned source points (in the 2D-case), are used). If no vector $w \in \rr^n$ satisfies the set of $(l-1)$ constraints 
\begin{equation} \label{overconstrained-proposition}
d_1(1+w^{\top}r_1)=\ldots=d_l(1+w^{\top}r_l)
\end{equation}
then the observer globally exponentially stabilizes $\tilde{X}=0$.
\item {\em Moving body}:\\
If the following two condition are satisfied:\\
${\bf (C_1)}$: there exist $\delta>0$~and~$\mu>0$ such that $\forall t \geq 0$:
\begin{equation} \label{condition1}
ZD^{\top}(\alpha)D(\alpha)Z^{\top}+\frac{1}{\delta} \int_t^{t+\delta} u(s)u^{\top}(s) ds >\mu I_d
\end{equation}
${\bf (C_2)}$: there exists $\nu>0$ such that $\forall t \geq 0$:\\ 
\begin{equation} \label{condition2}
\exists \tau \in [t,t+\delta]: |\dot{\bar{y}}(\tau)|>\nu
\end{equation}
then the observer globally exponentially stabilizes $\tilde{X}=0$.
\end{enumerate}
\end{lemma}
The proof of this lemma is given in Appendix \ref{proof-biased-range-measurements}.\\~\\
{\em Remarks}:\\
Concerning the first item of the lemma, one easily verifies that the constraints \eqref{overconstrained-proposition} have always a solution $w$ in the case of $(n+1)$ source points and that no solution "generically" exists in the case of $(n+2)$ and more source points. Two noticeable particular situations for which a solution exists independently of the number of source points are i) when all distances $d_i$ are equal ($w=0$ is then a solution), and ii) when all source points are located on the same half-side of a circular cone (in the 3D-case), or a conic sector (in the 2D-case), whose apex coincides with the body location ($w=\mu/sin(\theta)$, with $\mu$ the unit vector along the cone's axis and $\theta$ the cone's semi-angle, is then a solution). The exponential stability of the observer is thus not granted for these situations, as confirmed by simulation.\\
The second item of the lemma illustrates the interest of body motion in terms of observability and observer's performance, especially when less that $(n+2)$ source points are used. In particular, condition ${\bf (C_1)}$ is automatically satisfied in the 3D-case (resp. 2D-case) when at least four non-coplanar (resp. three non-aligned) source points are used and it can be satisfied with less source points when the body moves. As for condition ${\bf (C_2)}$, in the case of two and more source points, it is satisfied when $|u(t)|$ is regularly larger than some positive number. In the case of a single source point its satisfaction further requires that the body does not move on a sphere centred on the source point so that the distance between the body and the source point does not remain constant.  

\section{Simulations}\label{sec:simulations}
\begin{figure}
\centering
\subfigure[Trajectories of the body position and its estimates]{
\label{scenario1:a}
\includegraphics[scale=1.3]{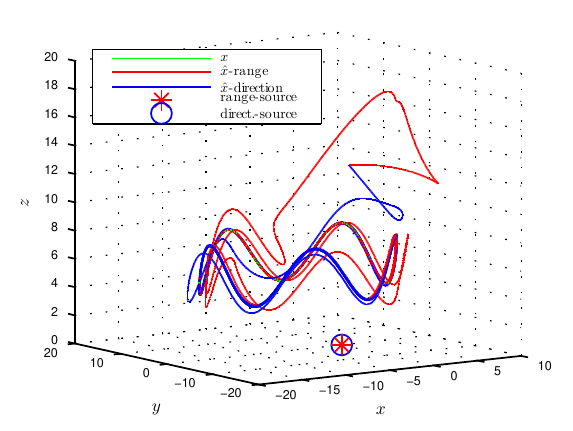}}
\caption{Scenario 1}
\label{scenario1}
\end{figure}

For these simulations in 3D-space we have considered three scenarios involving various body motions. Estimation of the body position is carried out from range and direction measurements with a minimal number of source points ensuring uniform observability. Since the conditions of observability are different in the two measurement cases, the number of source points may also be different. In all scenarios the body velocity measurement is corrupted by the constant bias $a=(0.33,\;0.66,\;0.99)^\top$ and initial state conditions are $x(0)=(5,\;0\;4)^\top$, $\hat{x}(0)=(4, \;6, \;12)^\top$ and $\hat{a}(0)=(0, \; 0,\; 0)^\top$. Riccati observers are calculated with $k(t)=1$, as for a Kalman filter, and the corresponding CRE are initialized with $P(0)=100 I_6$, when using direction measurements, and $P(0)=100 I_9$, when using range measurements. For each scenario, simulations are first carried out with noise-free measurements to validate theoretical exponential stability results, then with measurements corrupted by noise to illustrate the resulting (and inevitable) slight degradation of the observers following the transient phase when the estimation errors become small but no longer converge to zero. Concerning the body velocity $u$ we have used a Gaussian zero mean additive noise with a standard deviation equal to $0.1m/s$. As for the direction and range measurements, they are calculated from a body position corrupted by a Gaussian zero mean noise with standard deviation equal to $0.05m/s$. For the matrix $V$ (i.e. the state noise variance in the Kalman filtering terminology) involved in the CRE we have set $V=0.01 diag\{1,1,1,0,0,0\}+\epsilon_v I_6$, when using direction measurements, and $V=0.01 diag\{1,1,1,0,0,0,10,0,0\}+\epsilon_v I_9$, when using range measurements, with the small number $\epsilon_v$ set equal to $0.001$ to ensure that the matrix is positive definite. As for the matrix $Q$ (i.e. the inverse of the output noise variance in the Kalman filtering terminology) we have used $Q_{ii}=1.5 I_3, \; \forall i=1 \dots l$ (with $l$ the number of source points) and $Q=1.5 I_{l+1}$ respectively.
\\~\\
{\em Scenario 1}: The body moves along a Lissajous curve of equation
\[x(t)=(20 \cos t-15, \; 20 \sin t, \; -2 \cos t+6)^\top,\]
and a single source point located at the origin of the inertial frame is used for both direction and range measurements. One easily verifies that conditions for uniform observability are then satisfied in both cases. Figures \ref{scenario1:a}-\ref{scenario1:c} illustrate the performance of the two observers in the ideal noise-free case. More precisely, Figure \ref{scenario1:a} shows the location of the source point and the trajectories followed by the body position $x(t)$ and its estimate $\hat{x}(t)$, Figure \ref{scenario1:b} shows the convergence of the bias estimate $\hat{a}$ to the velocity bias $a$ and Figure \ref{scenario1:c} shows the evolution of the logarithms of the Lyapunov functions associated with the observers. The rate of exponential convergence to zero of the Lyapunov functions are given by the mean slopes of the curves. Asymptotic estimation errors in the case of noisy measurements are shown in Figures \ref{scenario1:d} and \ref{scenario1:e}.
\\~\\
{\em Scenario 2}: The body moves along a circular trajectory of equation 
\[x(t)=(20 \cos t-15, \; 20 \sin t, \; 4)^\top.\]
In this case a single source point, again taken as the origin of the inertial frame, suffices to ensure uniform observability in the direction measurement case, whereas a second source point has to be used in the range measurement case to ensure the satisfaction of this property. Figures \ref{scenario2:a}-\ref{scenario2:c} illustrate the performance of the two observers in the ideal noise-free case, and Figures \ref{scenario2:d}-\ref{scenario2:e} show asymptotic estimation errors in the case of noisy measurements.
\\~\\
{\em Scenario 3}: The body is motionless. Two source points are then needed in the direction measurement case to ensure uniform observability, whereas two other source points, non coplanar with them, are required in the range measurement case.  Figures \ref{scenario3:a}-\ref{scenario3:c} illustrate the performance of the two observers in the ideal noise-free case, and Figures \ref{scenario3:d}-\ref{scenario3:e} show asymptotic estimation errors in the case of noisy measurements. By comparison with the previous two scenarios the estimation errors are smaller. This is coherent with the increased number of source points that yields less noisy information in the average. 

\addtocounter{figure}{-1} 
\begin{figure}
\addtocounter{subfigure}{1} 
\centering 
\subfigure[Time evolution of the velocity bias estimation]{
\label{scenario1:b}
\includegraphics[scale=1.6]{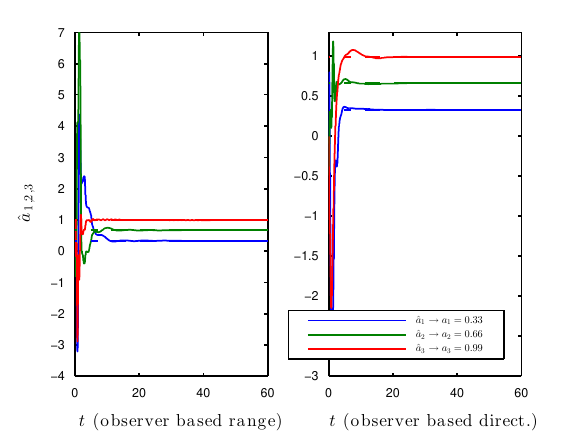}}
\caption{Scenario 1}
\end{figure}

\addtocounter{figure}{-1} 
\begin{figure}
\addtocounter{subfigure}{2} 
\centering 
\subfigure[Time evolution of the Lyapunov functions logarithms]{
\label{scenario1:c}
\includegraphics[scale=1.3]{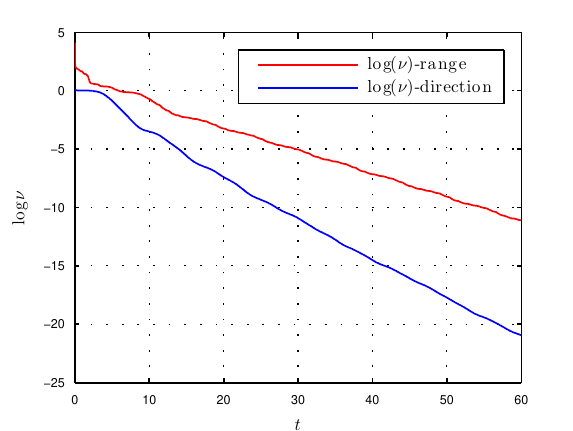}}
\caption{Scenario 1}
\end{figure}

\addtocounter{figure}{-1} 
\begin{figure}
\addtocounter{subfigure}{3} 
\centering 
\subfigure[Time evolution of velocity bias estimation errors in the case of noisy measurements]{
\label{scenario1:d}
\includegraphics[scale=1.6]{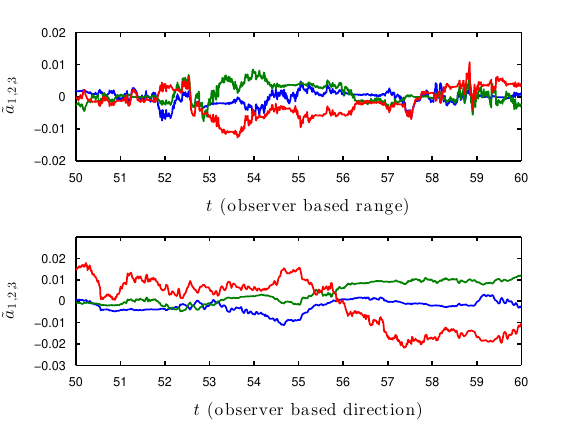}}
\caption{Scenario 1}
\end{figure}

\addtocounter{figure}{-1} 
\begin{figure}
\addtocounter{subfigure}{4} 
\centering 
\subfigure[Time evolution of position estimation errors in the case of noisy measurements]{
\label{scenario1:e}
\includegraphics[scale=1.6]{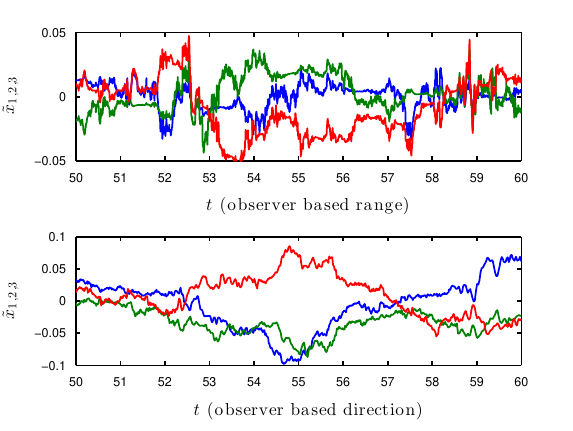}}
\caption{Scenario 1}
\end{figure}

\begin{figure}
\centering
\subfigure[Trajectories of the body position and its estimates]{
\label{scenario2:a}
\includegraphics[scale=1.3]{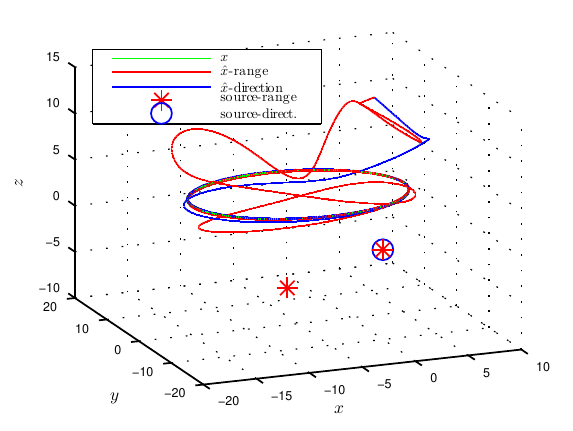}}
\caption{Scenario 2}
\label{scenario2}
\end{figure}

\addtocounter{figure}{-1} 
\begin{figure}
\addtocounter{subfigure}{1} 
\centering 
\subfigure[Time evolution of the velocity bias estimation]{
\label{scenario2:b}
\includegraphics[scale=1.6]{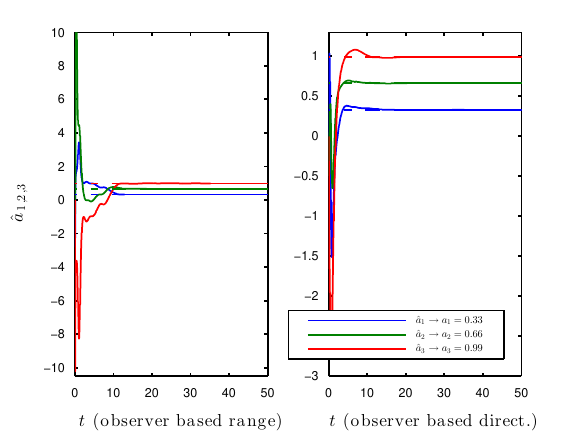}}
\caption{Scenario 2}
\end{figure}

\addtocounter{figure}{-1} 
\begin{figure}
\addtocounter{subfigure}{2} 
\centering 
\subfigure[Time evolution of the Lyapunov functions logarithms]{
\label{scenario2:c}
\includegraphics[scale=1.3]{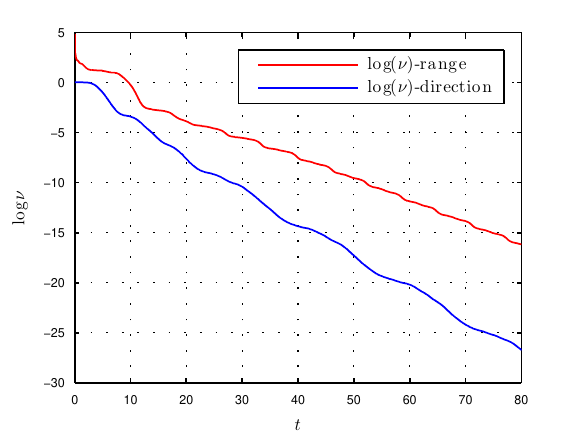}}
\caption{Scenario 2}
\end{figure}

\addtocounter{figure}{-1} 
\begin{figure}
\addtocounter{subfigure}{3} 
\centering 
\subfigure[Time evolution of velocity bias estimation errors in the case of noisy measurements]{
\label{scenario2:d}
\includegraphics[scale=1.6]{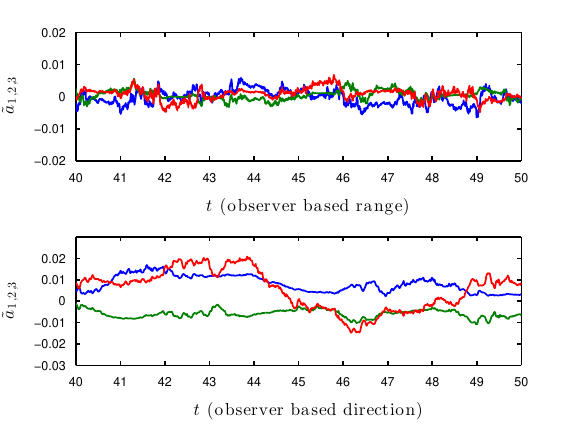}}
\caption{Scenario 2}
\end{figure}

\addtocounter{figure}{-1} 
\begin{figure}
\addtocounter{subfigure}{4} 
\centering 
\subfigure[Time evolution of position estimation errors in the case of noisy measurements]{
\label{scenario2:e}
\includegraphics[scale=1.6]{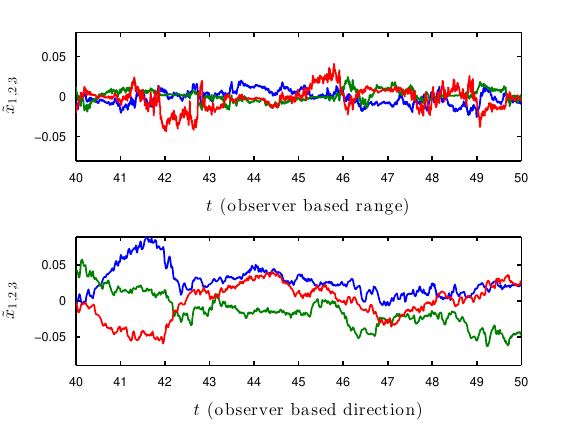}}
\caption{Scenario 2}
\end{figure}

\begin{figure}
\centering
\subfigure[Trajectories of the body position and its estimates]{
\label{scenario3:a}
\includegraphics[scale=1.3]{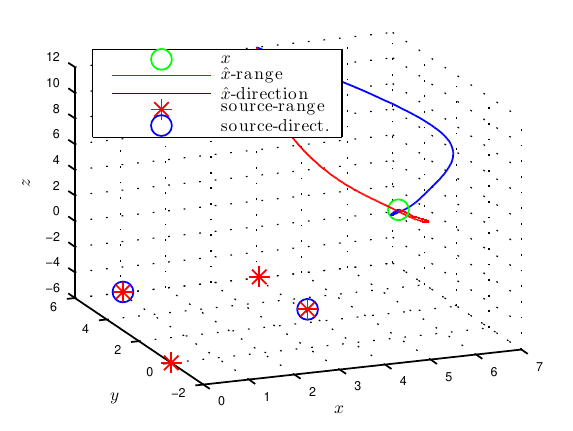}}
\caption{Scenario 3}
\label{scenario3}
\end{figure}

\addtocounter{figure}{-1} 
\begin{figure}
\addtocounter{subfigure}{1} 
\centering 
\subfigure[Time evolution of the velocity bias estimation]{
\label{scenario3:b}
\includegraphics[scale=1.6]{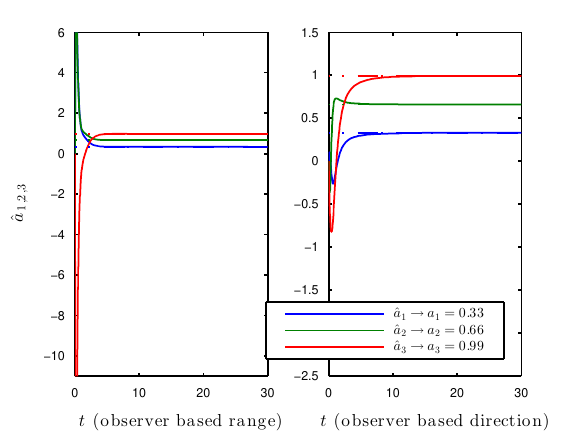}}
\caption{Scenario 3}
\end{figure}

\addtocounter{figure}{-1} 
\begin{figure}
\addtocounter{subfigure}{2} 
\centering 
\subfigure[Time evolution of the Lyapunov functions logarithms]{
\label{scenario3:c}
\includegraphics[scale=1.3]{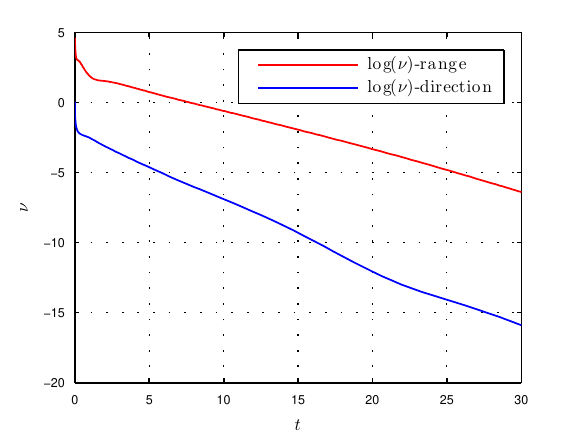}}
\caption{Scenario 3}
\end{figure}

\addtocounter{figure}{-1} 
\begin{figure}
\addtocounter{subfigure}{3} 
\centering 
\subfigure[Time evolution of velocity bias estimation errors in the case of noisy measurements]{
\label{scenario3:d}
\includegraphics[scale=1.6]{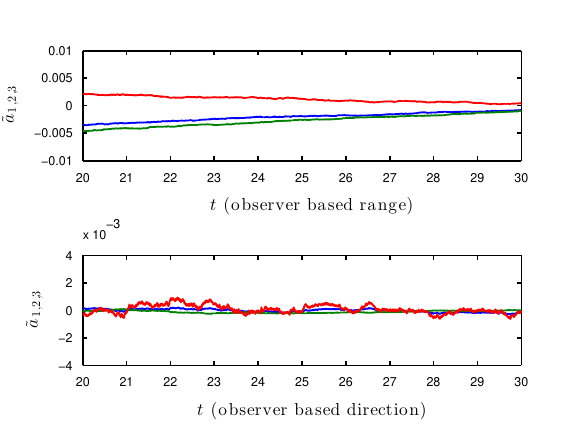}}
\caption{Scenario 3}
\end{figure}

\addtocounter{figure}{-1} 
\begin{figure}
\addtocounter{subfigure}{4} 
\centering 
\subfigure[Time evolution of position estimation errors in the case of noisy measurements]{
\label{scenario3:e}
\includegraphics[scale=1.6]{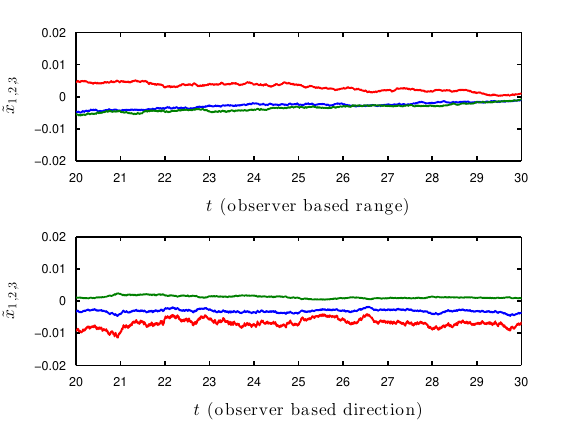}}
\caption{Scenario 3}
\end{figure}

\section{Concluding remarks}\label{sec:conclusions}
In this paper, Riccati observers for the estimation of a body position from either direction or range measurements and from the knowledge of the body velocity have been reviewed. Even when the body velocity is biased by an unknown constant vector, these observers ensure global exponential stability of zero estimation errors under uniform observability conditions that have been worked out in relation to the number of source points and the body motion. Clearly the set of such observers extends without difficulty to the case where the available information comes from the combination of direction measurements (associated with certain source points) with range measurements (associated with other source points). A logical prolongation of this work is the derivation of Riccati observers for the estimation of the complete body pose (position and orientation). However, due to the specific structure of the group of rotations, exact linearisation of the problem is then no longer possible and globally convex cost functions do not exist. As a consequence Riccati observers for pose estimation, and corresponding Extended Kalman Filters (EKF), have to be derived from linear approximations of the system state and output equations. This also implies that only local exponential stability of zero estimation errors can be achieved. An important complementary issue, also in the prolongation of the present work, is the characterisation of uniform observability conditions under which this latter property is granted. We foresee several other possible extensions. Let us just mention vision-based robotic applications involving the control of the body position from estimates provided by Riccati observers, and a deterministic approach to Simultaneous Localication and Mapping (SLAM) that could usefully complement existing studies on the subject. 

\section{Acknowledgments}
This work was supported by the ANR-ASTRID project SCAR ``Sensory
Control of Unmanned Aerial Vehicles'', the ANR-Equipex project ''Robotex''.

\appendix
\setcounter{section}{0}
\subsection{Proof of lemma \ref{existence}} \label{proof-existence}
Recall that, as long as $P(t)$ is defined and p.d., its trace is the sum of its eigenvalues. Accordingly, since the eigenvalues of $P^{-1}(t)$ are the inverse of the ones of $P(t)$, the trace of $P^{-1}(t)$ is the sum of the inverse of the eigenvalues of $P(t)$.
To prove that $P(t)$ is well-defined for $t \in [0,+\infty)$ and is p.d. it suffices to show that neither the trace of $P(t)$ nor the trace of $P^{-1}(t)$, which are initially positive (since $P(0)$ is p.d. by assumption), can tend to infinity in finite time. Indeed, this implies that none of the eigenvalues of $P(t)$ can either reach zero or tend to infinity in finite time. To this aim, it suffices to show that neither $tr(P(t)$ nor $tr(P^{-1}(t))$ can grow faster than exponentially, so that divergence in finite time is not possible.

Let us set $x=tr(P)$. In view of \eqref{dP}, and since $tr(P(t)C^{\top}(t)Q(t)C(t)P(t)) \geq 0$, one has
\[
\dot{x} \leq tr(AP)+tr(PA^{\top})+tr(V)
\]
Let $|A(t)|$ denote the spectral norm of $A(t)$. By assumption it is bounded by some positive number $k_a$. Similarly, $tr(V(t))$ is bounded by a positive number $v$. Since $P$ is p.s.d., $|tr(AP)|=|tr(PA^{\top})|\leq |A|tr(P)$ and the previous inequality yields
\[
\dot{x} \leq 2k_ax+v
\]
This inequality in turn implies that $x(t)\leq (x(0)+\frac{v}{2k_a})exp(2k_at)-\frac{v}{2k_a}$, $\forall t \geq 0$.

Similar arguments applied to $y=tr(P^{-1})$ yield
\[
\begin{array}{lll}
\dot{y}& \leq &|tr(P^{-1}A)|+|tr(A^{\top}P^{-1})+tr(C^{\top}QC)\\
~ & \leq & 2k_ay+\bar{\mu}_q
\end{array}
\]
with $\bar{\mu}_q$ denoting the supremum of $tr(C^{\top}(t)Q(t)C(t))$. Therefore, $y(t)\leq (y(0)+\frac{\bar{\mu}_q}{2k_a})exp(2k_at)-\frac{\bar{\mu}_q}{2k_a}$, $\forall t \geq 0$.\\
(end of proof)

\subsection{Determination of ultimate bounds for $p_m$ and $p_M$} \label{bounds}
\subsubsection{Ultimate lower bound of the smallest eigenvalue of $P(t)$ when $v_m>0$}
In practice the matrix $V(t)$ is usually chosen strictly positive so that the assumption of positivity on $v_m$ is little restrictive. Set, as in the previous appendix, $y(t)=tr(P^{-1}(t))=\sum_{i=1}^n \frac{1}{\lambda_i(t)}$, with $\{\lambda_i(t)\}_{i=1\ldots n}$ the set of eigenvalues of $P(t)$. The suprema of the spectral norm of $A(t)$ and of $tr(C^{\top}(t)Q(t)C(t))$ are again denoted as $k_a$ and $\bar{\mu}_q$ respectively. From \eqref{dPinverse} one has
\[
\dot{y}\leq 2k_ay+\bar{\mu}_q-tr(P^{-1}VP^{-1})
\]
with $tr(P^{-1}VP^{-1}) \geq \frac{v_m}{n}y^2$. This inequality implies that $y(t)$ is ultimately smaller than, or equal to, the largest (positive) root of the second degree equation $\frac{v_m}{n}y^2-2k_ay-\bar{\mu}_q=0$. More precisely
\[
\limsup_{t\rightarrow \infty}y(t) \leq \frac{n k_a}{v_m}\big( 1+(1+\frac{\bar{\mu}_qv_m}{n k_a^2})^{0.5} \big)
\]
Since $1/\lambda_{min}(P(t)) \leq y(t)$ the previous inequality yields
\begin{equation} \label{ultimate_lower}
\liminf_{t\rightarrow \infty}\lambda_{min}(P(t)) \geq \frac{v_m}{n k_a}\big( 1+(1+\frac{\bar{\mu}_qv_m}{n k_a^2})^{0.5}\big)^{-1}
\end{equation}

\subsubsection{Ultimate upper bound of the largest eigenvalue of $P(t)$ when $W_Q(t,t+\delta)\geq \mu I_d>0$} \label{bounds-estimation}
Recall that $W_Q$ is the Riccati observability Grammian defined in \eqref{Wq}. We use the expression of the upper bound of $P(t)$ derived in \cite{PRV2001}
{\small
\[
P(t)\leq \delta W_Q^{-1}(t,t+\delta)+\delta^2 W_Q^{-1}(t,t+\delta)I_2(t,t+\delta)W_Q^{-1}(t,t+\delta)
\]
}
with $I_2(t,t+\delta)$ a positive matrix-valued function which, using the inequality $|\Phi(t,s)|\leq exp(k_a|t-s|)I_d$, is upper bounded by $\frac{exp(6k_a\delta)\bar{\mu}_q^2\delta^3v_M}{3}I_d$. Therefore, when $W_Q(t,t+\delta)\geq \mu I_d>0$ one deduces that
\begin{equation} \label{upperbound}
\limsup_{t\rightarrow \infty}\lambda_{max}(P(t)) \leq \frac{1}{\mu \delta}+\frac{1}{3}\big(\frac{\bar{\mu}_q}{\mu}\big)^2 exp(6k_a\delta)\delta v_M
\end{equation}

Relations \eqref{ultimate_lower} and \eqref{upperbound} can in turn be used to estimate an ultimate lower bound of $\frac{p_m^2}{p_M}v_m$, i.e. an estimate of the lower bound pointed out in \eqref{dV} of the exponential rate of convergence associated with a Riccati observer.

\subsection{Proof of lemma \ref{extension}} \label{proof-extension}
For the sake of simplifying the reading of the proof by avoiding non-essential technicalities, we set $Q=k_q I_d$ with $k_q>0$.
Let us proceed by contradiction and assume that the lemma's conclusion is wrong, i.e.
\[
\forall \epsilon, \forall \bar{\delta}>0, \exists t \geq 0~:~W_Q(t,t+\bar{\delta})<\epsilon I_d
\]
Consider a sequence $\{\epsilon_p\}_{p\in\nn}$ of positive numbers converging to zero, and an arbitrary positive number $\bar{\delta}$. From the previous assertion there must exist a sequence of time-instants $\{t_p\}_{p\in\nn}$ and a sequence $\{x_p\}_{p\in\nn}$ with $x_p \in S^{n-1}$ (i.e. $|x_p|=1$) such that $\forall p \in \nn~:~x_p^{\top}W_Q(t_p,t_p+\bar{\delta})x_p<\epsilon_p$. Since $S^{n-1}$ is a compact set there exists a sub-sequence of $\{x_p\}_{p\in\nn}$ which converges to a limit $\bar{x} \in S^{n-1}$. Therefore
\[
\lim_{p\rightarrow \infty} \bar{x}^{\top}W_Q(t_p,t_p+\bar{\delta})\bar{x}=0
\]
Using $C=\Pi_{y(t)}\bar{C}$ and $\Phi(t,s)=exp(A(t-s))$ in the definition \eqref{Wq} of $W_Q$, the above equality is equivalent to
\[
\lim_{p\rightarrow \infty} \int_0^{\bar{\delta}} |\Pi_{y(t_p+s)}\bar{C} exp(As)\bar{x}|^2ds=0
\]
which in turn implies
\begin{equation} \label{lim}
\lim_{p\rightarrow \infty} \int_{\bar{\delta}-\delta}^{\bar{\delta}} |\Pi_{y(t_p+s)}\bar{C} exp(As)\bar{x}|^2ds=0
\end{equation}
provided that $\bar{\delta} \geq \delta$.
Consider now the following technical result whose proof is given at the end of the present appendix
\begin{lemma} \label{technical}
Assume that the eigenvalues of the matrix $A$ are all real, then, given $\bar{x}\in S^{n-1}$, there exist $r \geq 0$, $\lambda \in \rr$, and $z\in \rr^m-\{0 \}$ such that $\frac{\bar{C} exp(At)\bar{x}}{t^r exp(\lambda t)}=z+\eta(t)$ with $\lim_{t\rightarrow +\infty}\eta(t)=0$.
\end{lemma}
In view of this result, setting $\bar{z}=z/|z| \in S^{m-1}$, and choosing $\bar{\delta}$ large enough so that $sup_{s\in [\bar{\delta}-\delta,\bar{\delta}]}|\eta(s)|<\sqrt{\frac{\epsilon}{2}}|z|$ one deduces that
\[
\begin{array}{l}
\frac{1}{\gamma^2|z|^2}\int_{\bar{\delta}-\delta}^{\bar{\delta}} |\Pi_{y(t_p+s)}\bar{C} exp(As)\bar{x}|^2ds \\ \geq
\int_{\bar{\delta}-\delta}^{\bar{\delta}} |\Pi_{y(t_p+s)}(\bar{z}+\frac{\eta(s)}{|z|})|^2ds \\
\geq
\int_{t_p+\bar{\delta}-\delta}^{t_p+\bar{\delta}} |\Pi_{y(s)}\bar{z}|^2ds-\int_0^{\bar{\delta}} |\Pi_{y(t_p+s)}\frac{\eta(s)}{|z|})|^2ds \\
\geq~\delta\epsilon - \delta\epsilon/2~ (=\delta\epsilon/2)
\end{array}
\]
with $\gamma=inf_{s\in [\bar{\delta}-\delta,\bar{\delta}]}(s^r exp(\lambda s))>0$. The p.e. condition \eqref{persistent} is used in the last inequality. Therefore
\[
\int_{\bar{\delta}-\delta}^{\bar{\delta}} |\Pi_{y(t_p+s)}\bar{C} exp(As)\bar{x}|^2ds \geq \gamma^2|z|^2\frac{\epsilon}{2}>0
\]
Since this latter inequality holds true for any $t_p$, it contradicts \eqref{lim} and the initial assumption according to which the result of the lemma is not true.

It only remains to prove the technical Lemma \ref{technical}. From Cayley-Hamilton's theorem, one has
$exp(At)=\sum_{i=0}^{n-1}\alpha_i(t)A^i$ with $\alpha_i(t)=\sum_{k=1}^d(\sum_{j=0}^{l_k-1}a_{ij}t^j)exp(\lambda_kt)$, $\lambda_k$ a (real) eigenvalue of $A$, $a_{ij}\in\rr$, $d\leq n$ the number of distinct eigenvalues, and $l_k$ the multiplicity of $\lambda_k$. Therefore
\[
\begin{array}{lll}
\bar{C}exp(At)\bar{x} & = & \bar{C} \sum_{i=0}^{n-1}\alpha_i(t)A^i \bar{x}
= \sum_{i=0}^{n-1}\alpha_i(t)\bar{C}A^i \bar{x}\\
~ & = & \sum_{i=0}^{n-1}\alpha_i(t)z_i
\end{array}
\]
with $\bar{\bar{z}}=\left[ \begin{array}{c}z_1\\ \vdots \\z_n \end{array} \right]:={\cal{O}} \bar{x}$ and ${\cal{O}}=\left[ \begin{array}{c} \bar{C} \\ \bar{C}A \\ \vdots \\ \bar{C}A^{n-1} \end{array} \right]$ the Kalman observability matrix whose rank is, by assumption, equal to $n$. This latter assumption in turn implies that the vector $\bar{\bar{z}}$ is different from zero, and thus that at least one of the $z_i$ components of this vector is different from zero.
The previous sum can also be arranged as follows
\[
\sum_{i=0}^{n-1}\alpha_i(t)z_i=\sum_{k,j}v_{k,j}(t)\bar{z}_{k,j}~~~\bar{z}_{k,j}\in\rr^m
\]
with $v_{k,j}(t)=t^{r_{k,j}}exp(\lambda_kt)$, $k\in [1,\ldots,n]$, $r_{k,j}\in [0,\ldots,n-1]$. We note that at least one of the vectors $\bar{z}_{k,j}$ must be different from zero, due to the observability assumption and the full rank of ${\cal{O}}$. Consider the largest (less negative, or most positive) root $\lambda_k$ for which $\bar{z}_{k,j}$ is different from zero, and the largest power $r_{k,j}$ that goes with such a vector. Denote this root as $\lambda$ and this power as $r$, set $v(t):=t^r exp(\lambda t)$, and denote the corresponding vector $\bar{z}_{k,j}$ as $z$ ($\neq 0$). The dominating coefficient in the development of $\bar{C}exp(At)\bar{x}$, when $t$ tends to infinity, is thus $v(t)$ and one has
$\lim_{t\rightarrow \infty}\frac{\bar{C}exp(At)\bar{x}}{v(t)}=z$. This latter property can also be written as $\frac{\bar{C}exp(At)\bar{x}}{v(t)}=z+\eta(t)$ with $\lim_{t\rightarrow \infty}\eta(t)=0$.

\subsection{Proof of lemma \ref{excitation_range_unbiased}} \label{proof_excitation_range_unbiased}
Recalling that the positivity of the observability Grammian $W$ yields the positivity of the Riccati observability Grammian $W_Q$ when $Q(t)\geq \epsilon I_d >0$, one only has to show --according to Lemma \ref{scandaroli}-- the existence of an adequate matrix-valued function $M(.)$ that satisfies \eqref{M} for some positive numbers $\bar{\delta}$ and $\bar{\mu}$.

For the system under consideration one has $N_0=C=[0_{1\times n}~ 1]$ and $N_1(t)=CA(t)=[u^{\top}(t)~ 0]$. Define
\[
M(t):=\left[ \begin{array}{c} N_0 \\ N_1(t)\end{array} \right]=\left[ \begin{array}{cc} 0_{1 \times n} & 1 \\ u^{\top}(t) & 0 \end{array} \right]
\]
and consider an arbitrary vector $b\in S^n$. Then $M(t)b=\left[ \begin{array}{c} b_2 \\ u^{\top}(t)b_1 \end{array} \right]$ with $b=[b_1^{\top},b_2]^{\top}$. Therefore $|M(t)b|^2=b_1^{\top}u(t)u^{\top}(t)b_1+b_2^2$. Define
$\gamma(t):= \frac{1}{\delta}\int_t^{t+\delta}|M(s)b|^2ds$. Using the fact that $b_2^2=1-|b_1|^2$ one has
\[
\gamma(t)=(1-|b_1|^2)+\frac{1}{\delta}\int_t^{t+\delta}b_1^{\top}u(s)u^{\top}(s)b_1ds
\]
with $0\leq |b_1| \leq 1$. There are two possible cases: either $b_1=0$ or $b_1\neq 0$. In the first case one obtains $\gamma(t)=1$. In the second case, setting $\bar{b}_1:=b_1/|b_1| \in S^{n-1}$, one obtains $\gamma(t)=(1-|b_1|^2)+
\frac{|b_1|^2}{\delta}\int_t^{t+\delta}\bar{b}_1^{\top}u(s)u^{\top}(s)\bar{b}_1ds \geq (1-|b_1|^2)+|b_1|^2\mu \geq \inf(1,\mu)$. Therefore  $\gamma(t) \geq \bar{\mu}$ with  $\bar{\mu}=\inf(1,\mu)$. Since the last inequality holds for any $b\in S^n$, \eqref{M} holds true.

\subsection{Proof of lemma \ref{excitation_range_biased}} \label{proof_excitation_range_biased}
As in the unbiased case we show the existence of a matrix-valued function $M(.)$ that satisfies \eqref{M} for some positive numbers $\bar{\delta}$ and $\bar{\mu}$.
For the system under consideration one has $N_0=C=[0_{1\times n}~0_{1\times n}~ 1~0~0]$,  $N_1(t)=CA(t)=[u^{\top}(t)~0_{1\times n}~0~1~0]$, and $N_2(t)=N_1(t)A(t)+\dot{N}_1(t)=[\dot{u}^{\top}(t)~2u^{\top}(t)~0~0~1]$. Define
\[
M(t):=\left[ \begin{array}{c} N_0 \\ N_1(t)\\ N_2(t)\end{array} \right]
\]
and consider an arbitrary vector $b=[b_1^{\top},b_2^{\top},b_3^{\top}]^{\top}\in S^{2n+2}$, with $b_{1,2,3}$ sub-vectors of dimensions $n$, $n$, and $3$ respectively. Then $M(t)b=\left[ \begin{array}{c} 0 \\ u^{\top}(t)b_1 \\ \dot{u}^{\top}(t)b_1+2u^{\top}(t)b_2 \end{array} \right]+b_3$ and $|M(t)b|^2=b_{3,1}^2+(u^{\top}(t)b_1+b_{3,2})^2+(\dot{u}^{\top}(t)b_1+2u^{\top}(t)b_2+b_{3,3})^2$, with $b_{3,i}$ denoting the $ith$ component of $b_3$. Define
$\gamma(t):= \frac{1}{\delta}\int_t^{t+\delta}|M(s)b|^2ds$ and let us make a proof by contradiction by assuming that the condition \eqref{M} is not satisfied. In this case there exists a sequence $\{t_p\}$ and a vector $b \in S^{2n+2}$ such that $\lim_{p\rightarrow +\infty}\gamma(t_p)=0$. This in turn implies that $b_{3,1}=0$ and also
\begin{equation} \label{limit1}
\lim_{p\rightarrow +\infty} \int_{t_p}^{t_p+\delta}(u^{\top}(s)b_1+b_{3,2})^2ds=0
\end{equation}
and
\begin{equation} \label{limit2}
\lim_{p\rightarrow +\infty} \int_{t_p}^{t_p+\delta}(\dot{u}^{\top}(s)b_1+2u^{\top}(s)b_2+b_{3,3})^2ds=0
\end{equation}
Using the assumed boundedness of $\dot{u}(t)$ the first of these two limits yields $\lim_{p\rightarrow +\infty}u^{\top}(t_p+s)b_1=-b_{3,2}$, $\forall s \in (0,\delta)$. Using now the assumed boundedness of $\ddot{u}(t)$ this in turn implies that $\lim_{p\rightarrow +\infty}\dot{u}^{\top}(t_p+s)b_1=0$, $\forall s \in (0,\delta)$.
From \eqref{limit2} one deduces that $\lim_{p\rightarrow +\infty}(\dot{u}^{\top}(t_p+s)b_1+2u^{\top}(t_p+s)b_2+b_{3,3})=0$, $\forall s \in (0,\delta)$ and, subsequently, that $\lim_{p\rightarrow +\infty}u^{\top}(t_p+s)b_2=-b_{3,3}/2$, $\forall s \in (0,\delta)$. Using the assumed boundedness of $\ddot{u}(t)$ this in turn implies that $\lim_{p\rightarrow +\infty}\dot{u}^{\top}(t_p+s)b_2=0$,
$\forall s \in (0,\delta)$. If either $b_1$ or $b_2$ is different from zero one reaches a contradiction with \eqref{pedu}. Therefore $b_1=b_2=0$. But then, from what precedes, $b_{3,2}=b_{3,3}=0$ so that $b=0$. This is not possible since $b\in S^{2n+2}$.

\subsection{Proof of lemma \ref{proposition}} \label{proof-biased-range-measurements}
Define
\[
M(t):=\left[ \begin{array}{c} N_0(t) \\ \bar{N}_1(t) \end{array} \right]
\]
with $N_0(t)=C(t)$, $N_1(t)=C(t)A(t)$, and $\bar{N}_1(t)$ the first line of $N_1(t)$, i.e.
\begin{equation} \label{M}
M(t)=\left[ \begin{array}{ccc}
0_{1 \times n} & 1 & \alpha^{\top}\bar{y}(t) \\
D(\alpha)Z^{\top} & 0 & -D(\alpha)\bar{y}(t) \\
u(t)^{\top} & 0 & 0 
\end{array} \right]
\end{equation}
We make a proof by contradiction of the first result by assuming that a uniform observability condition yielding the uniform exponential stability of the observer is not satisfied when \eqref{overconstrained-proposition} does not have a solution. More precisely we assume that
\begin{equation} \label{assumption1}
\forall \epsilon \geq 0,~\forall v \in S^{n+1},~\exists t\geq 0: \int_t^{t+\delta}|M(s)v|^2ds <\epsilon
\end{equation}
with $n=3$ (the 3D case) or $n=2$ (the 2D case). Let $v_1 \in \rr^n$, $v_2 \in \rr$ and $v_3 \in \rr$ denote the components of the unit vector $v$. In this case $u(t)\equiv 0$ (motionless body) so that $M$ is a constant matrix and 
\[
Mv=\left[ \begin{array}{c}
v_2+\alpha^{\top}\bar{y}v_3\\
D(\alpha)Z^{\top}v_1-D(\alpha)\bar{y}v_3\\
0
\end{array} \right]
\]
Assume that $v_3=0$. Then 
\[
|M(s)v|^2 = v_2^2+v_1^{\top}ZD^{\top}(\alpha)D(\alpha)Z^{\top}v_1
\]
with $ZD^{\top}(\alpha)D(\alpha)Z^{\top}$ a p.d. matrix by assumption. Therefore $v^{\top}M^{\top}Mv \geq inf(1,\lambda_m)$ with $\lambda_m$ the smallest (strictly positive) singular value of $ZD^{\top}(\alpha)D(\alpha)Z^{\top}$. Since this contradicts \eqref{assumption1} one deduces that $v_3\neq 0$. Now, since $Mv$ is a constant vector, the satisfaction of \eqref{assumption1} implies the existence of a unit vector $v$ such that $Mv=0$. This in turn implies that $v_2=-\alpha^{\top}\bar{y}v_3$ and $D(\alpha)Z^{\top}v_1-D(\alpha)\bar{y}v_3=0$. Using the fact that $D(\alpha)=\xi\alpha^{\top}-I_{l \times l}$ the first of these equalities yields $D(\alpha)\bar{y}v_3=-\xi v_2-\bar{y}v_3$. Since $\bar{y}_i=d_i+b$, substracting the $ith$ line from the first line of the left member of the second equality yields $(z_i-z_1)^{\top}v_1=(d_i-d_1)v_3$ ($i=2,\ldots,l$). Therefore
\begin{equation} \label{overconstrained}
\left[ \begin{array}{c}
(z_2-z_1)^{\top} \\ \vdots \\(z_l-z_1)^{\top}
\end{array} \right] \bar{v}_1
=
\left[ \begin{array}{c} 
d_2-d_1 \\ \vdots \\ d_l-d_1
\end{array} \right]
\end{equation}
with $\bar{v}_1=v_1/v_3$. Since $z_i-z_1=d_1r_1-d_ir_i$ the previous equation is the same as $(d_1r_1^{\top}-d_ir_i^{\top})\bar{v}_1=d_i-d_1$ for $i=2,\ldots,l$, which in turn is the same as \eqref{overconstrained-proposition} with $w=\bar{v}_1$. Since this equation has no solution by assumption a contradiction is reached and the result follows.\\
We now prove the second result of the lemma.

Define $\gamma(v,t):=\int_t^{t+\delta}|M(s)v|^2ds$ and assume that \eqref{assumption1} holds true. Then there exists a sequence $\{t_p\}$ and a unit vector $v$ such that $\lim_{p \rightarrow +\infty}\gamma(v,t_p)=0$. This in turn implies that
\[
\lim_{p \rightarrow +\infty} \int_{t_p}^{t_p+\delta} (v_2+\alpha^{\top}\bar{y}(s)v_3)^2ds=0
\]
and
\[
\lim_{p \rightarrow +\infty} \int_{t_p}^{t_p+\delta} |D(\alpha)Z^{\top}v_1-D(\alpha)\bar{y}(s)v_3|^2ds=0
\]
and also 
\[
\lim_{p \rightarrow +\infty} \int_{t_p}^{t_p+\delta} |u^{\top}(s)v_1|^2ds=0
\]
Using the assumed boundedness of $\dot{u}(t)$, and thus of $\dot{\bar{y}}(t)$, the first of these equalities yields $\lim_{p \rightarrow +\infty} (\alpha^{\top}\bar{y}(t_p+s)v_3)=-v_2$, $\forall s \in (0,\delta)$. Using the assumed boundedness $\ddot{u}(t)$ this in turn implies that $\lim_{p \rightarrow +\infty} \alpha^{\top}\dot{\bar{y}}(t_p+s)v_3=0$, $\forall s \in (0,\delta)$. Using similar arguments for the second equality one deduces that $\lim_{p \rightarrow +\infty}D(\alpha)\dot{\bar{y}}(t_p+s)v_3=0$, $\forall s \in (0,\delta)$. Since $D(\alpha)=\xi \alpha^{\top}-I_{l\times l}$ one obtains that $\lim_{p \rightarrow +\infty}\dot{\bar{y}}(t_p+s)v_3=\xi\lim_{p \rightarrow +\infty}\alpha^{\top}\dot{\bar{y}}(t_p+s)v_3=0$, $\forall s \in (0,\delta)$. From condition ${\bf (C_2)}$ this in turn implies that $v_3=0$ and, subsequently that $v_2=0$. Combining the second and third equalities then yields
\[
\lim_{p \rightarrow +\infty}v_1^{\top}(ZD^{\top}(\alpha)D(\alpha)Z^{\top}+\frac{1}{\delta} \int_{t_p}^{t_p+\delta} u(s)u(s)^{\top} ds)v_1=0
\]
which, in view of condition ${\bf (C_1)}$, implies that $v_1=0$. Therefore $v=0$ and this contradicts the assumption according to which $v$ is a unit vector.
\bibliographystyle{IEEEtran}
\bibliography{bibfile3}

\end{document}